\documentclass[a4paper,11pt]{article}
\pdfoutput=1

\usepackage{jcappub}
\usepackage{amsmath}
\usepackage{amsfonts}
\usepackage{amssymb}
\usepackage{mathtools}
\usepackage{graphics}
\usepackage{tabularx}
\usepackage[export]{adjustbox}
\usepackage{multirow}
\usepackage{citesort}
\usepackage{graphicx}
\usepackage{url}
\usepackage{soul}
\usepackage{bm}
\usepackage[dvipsnames]{xcolor}
\usepackage[utf8]{inputenc}
\usepackage[normalem]{ulem}
\usepackage{hyperref}

\title{Perturbing Fast Neutrino Flavor Conversion}

\author[a]{Marie Cornelius,}
\author[a]{Shashank Shalgar,}
\author[a]{and Irene Tamborra}

\affiliation[a]{Niels Bohr International Academy and DARK, Niels Bohr Institute, University of Copenhagen, Blegdamsvej 17, 2100, Copenhagen, Denmark}

\emailAdd{marie.cornelius@nbi.ku.dk}
\emailAdd{shashank.shalgar@nbi.ku.dk}
\emailAdd{tamborra@nbi.ku.dk}

\abstract{The flavor evolution of neutrinos in dense astrophysical sources, such as core-collapse supernovae or compact binary mergers, is non-linear due to the coherent forward scattering of neutrinos among themselves. Recent work in this context has been addressed to figure out whether flavor equipartition could be a generic flavor outcome of fast flavor conversion. We investigate the flavor conversion physics injecting random perturbations in the neutrino field in two simulation setups: 1.~a spherically symmetric simulation shell without periodic boundaries, with angular distributions evolving dynamically thanks to non-forward scatterings of neutrinos with the background medium, and neutrino advection; 2.~a periodic simulation shell, with angular distributions of neutrinos defined a priori and neutrino advection. We find that, independent of the exact initial flavor configuration and type of perturbations, flavor equipartition is generally achieved in the system with periodic boundaries; in this case, perturbations aid the diffusion of flavor structures to smaller and smaller scales. However, flavor equipartition is not a general outcome in the simulation shell without periodic boundaries, where the inhomogeneities induced perturbing the neutrino field affect the flavor evolution, but do not facilitate the diffusion of flavor waves. This work highlights the importance of the choice of the simulation boundary conditions in the exploration of fast flavor conversion physics.
}

\begin{document}
\maketitle

\section{Introduction}

Despite their weakly interacting nature, neutrinos are crucial  in core-collapse supernovae and neutron star mergers~\cite{Burrows:2020qrp,Janka:2022krt,Janka:2016fox}.
In the core of these sources, the neutrino number density is so large that the coherent forward scattering of neutrinos off other neutrinos has a significant role, leading to fast flavor conversion~\cite{Sawyer:2005jk,Sawyer:2008zs,Sawyer:2015dsa}. The latter has shown to have a very rich phenomenology~\cite{Tamborra:2020cul,Chakraborty:2016yeg,Richers:2022zug}, which we are still far from fully grasping. While our understanding of the asymptotic flavor distribution is still under development, recent work based on  parametric modeling of the fast flavor conversion outcome points towards a non-trivial impact of fast conversion on neutrino-driven explosions of core-collapse supernovae~\cite{Ehring:2023lcd,Ehring:2023abs,Nagakura:2023mhr} as well as on the ejecta properties and disk cooling rate of neutron star merger remnants~\cite{Wu:2017drk,George:2020veu,Just:2022flt,Li:2021vqj}.

The flavor conversion phenomenology depends on the shape of the angular distributions of neutrinos of different flavors, see e.g.~Refs.~\cite{Tamborra:2020cul,Chakraborty:2016yeg}. In particular, a necessary condition to trigger fast flavor conversion is the presence of a crossing in the angular distributions of the number of neutrinos and antineutrinos of a certain flavor (so-called neutrino flavor lepton number crossing)~\cite{Izaguirre:2016gsx,Morinaga:2021vmc,Padilla-Gay:2021haz}. Within an idealized neutrino system, pairwise conversion of neutrinos could take place due to the existence of a neutrino flavor lepton number crossing, even in the limit of vanishing mass differences among the neutrino mass eigenstates. However, the non-trivial interplay among flavor conversion, non-forward collisions with the background medium, and advection strongly affects the flavor conversion history~\cite{Shalgar:2022rjj,Shalgar:2022lvv,Kato:2023dcw,Nagakura:2022qko,Shalgar:2019qwg,Padilla-Gay:2020uxa,Shalgar:2020wcx,Hansen:2022xza,Martin:2021xyl}. The fact that neutrinos exhibit a flavor-dependent spectral energy distribution, are not degenerate in mass, and exist in three different flavors, further impacts the final flavor configuration~\cite{Shalgar:2020xns,Shalgar:2021wlj,Capozzi:2022dtr,Capozzi:2020kge,Chakraborty:2019wxe}.
More recently, it has been shown that flavor instabilities could also be generated through collisions of neutrinos with the background medium~\cite{Johns:2021qby}; a phenomenon that still remains to be understood, especially for what concerns its astrophysical implications~\cite{Shalgar:2023aca,Xiong:2022vsy,Padilla-Gay:2022wck,Liu:2023vtz}.

The full solution of the neutrino quantum kinetics in seven dimensions has not been tackled yet because of the numerical challenges that this problem entails~\cite{Tamborra:2020cul,Richers:2022zug}. However, in order to assess the impact of flavor conversion on the source physics, the asymptotic flavor configuration needs to understood~\cite{Zaizen:2023ihz,Zaizen:2022cik,Nagakura:2022kic,Xiong:2023vcm}. In systems with periodic boundary conditions, flavor equipartition (at least on one side of the angular distribution with respect to the neutrino flavor lepton number crossing) has been reported as a likely outcome~\cite{Grohs:2022fyq,Richers:2021xtf,Richers:2022bkd,Bhattacharyya:2020jpj,Bhattacharyya:2022eed,Xiong:2023vcm,Wu:2021uvt,Martin:2019gxb,Martin:2021xyl,Abbar:2021lmm,Duan:2021woc,Zaizen:2023ihz,Zaizen:2022cik}, but flavor equipartition seems to be just one of the possible final flavor configurations once the assumption of periodic boundaries is relaxed~\cite{Shalgar:2022rjj,Shalgar:2022lvv,Padilla-Gay:2022wck,Nagakura:2022qko,Nagakura:2022kic,Nagakura:2022xwe,Sigl:2021tmj}.

The asymptotic flavor configuration could also be affected by fluctuations of the neutrino field linked to the source dynamics, see e.g.~Refs.~\cite{Tamborra:2014aua,Tamborra:2014hga,Walk:2019miz,Shibagaki:2020ksk,Takiwaki:2017tpe,Nagakura:2021lma}.
For example, employing periodic boundary conditions and perturbations extended in space, because of flavor instabilities, Ref.~\cite{Martin:2019gxb} found that small perturbations in the initial flavor distribution otherwise uniform in space develop into fast oscillation waves that can propagate and diffuse in space. While it has been shown that perturbations in the off-diagonal terms of the density matrices of neutrinos and antineutrinos speed up the evolution of the neutrino system to the non-linear phase~\cite{Grohs:2022fyq,Richers:2021xtf,Bhattacharyya:2020jpj,Bhattacharyya:2022eed}, the role of perturbations of the neutrino field on the quasi-steady state configuration remains poorly understood.

This paper aims to investigate whether space and time dependent fluctuations of the neutrino field affect the quasi-steady state configuration achieved by neutrinos and whether they aid the achievement of flavor equipartition. To this purpose, we solve the neutrino quantum kinetic equations in simulation shells with periodic and non-periodic boundary conditions to explore whether the choice of the boundary conditions affects the quasi-steady state flavor configuration.

Our work is organized as follows. In Sec.~\ref{Sec:QKE}, we introduce the neutrino equations of motion and the simulation setup without and with periodic boundary conditions. Section~\ref{Sec:no_perts} presents our results on the quasi-steady state neutrino configuration without any perturbation in the neutrino field. The model adopted to perturb the neutrino field and the impact of such perturbations on the neutrino flavor evolution is explored in Sec.~\ref{Sec:impact_of_perts} for the simulation shells with and without periodic boundaries. Our findings are summarized and discussed in Sec.~\ref{Sec:discussion}. Furthermore, in Appendix \ref{App:other_ang_dist}, we report results on the asymptotic flavor configuration obtained adopting a different set of neutrino angular distributions for the simulation shell with periodic boundaries; in Appendix \ref{App:radial_perts}, we investigate the impact on the flavor conversion physics of radial perturbations in the neutrino density.

\section{Neutrino quantum kinetics}
\label{Sec:QKE}

In this section, we introduce the neutrino equations of motion as well as the simulation setup. 

\subsection{Simulation shell without periodic boundaries}

For the sake of simplicity, we work in the two-flavor basis ($\nu_e$, $\nu_x$).
Assuming that neutrinos are mono-energetic, the neutrino and antineutrino flavor field can be modeled relying on $2 \times 2$ density matrices in the flavor space, $\rho(r,\cos\theta,t)$ and $\bar{\rho}(r,\cos\theta,t)$, respectively.
The diagonal elements of the density matrix, $\rho_{ii}$ (with $i=e, x$), represent the occupation number of neutrinos of a given flavor; instead, the off-diagonal elements, $\rho_{ij}$, track the flavor coherence. 
At the time $t$, the neutrino field is characterized by the radial coordinate $r$ and polar angle $\theta \equiv \theta(r)$. 

The quantum kinetic equations describing the flavor evolution of neutrinos and antineutrinos, respectively, are~\cite{Sigl:1993ctk}:
\begin{align}
    i\left(\frac{\partial}{\partial t} + \Vec{v}\cdot \vec\nabla \right) \rho(r,\cos\theta,t) &= \left[H,\rho(r,\cos\theta,t)\right] + i\mathcal{C} \label{Eq:QKEs1}\ ,\\
    i\left(\frac{\partial}{\partial t} + \Vec{v}\cdot \vec\nabla \right) \bar{\rho}(r,\cos\theta,t) &= \left[\bar{H},\bar{\rho}(r,\cos\theta,t)\right] + i\bar{\mathcal{C}}\ . \label{Eq:QKEs2}
\end{align}
The advective term in the left-hand side is
\begin{equation}
    \Vec{v}\cdot \vec\nabla \rho(r,\cos\theta,t) = 
    \frac{\partial \rho(r,\cos\theta,t)}{\partial \cos{\theta}}\frac{\sin^2\theta}{r} + \cos{\theta}\frac{\partial \rho(r,\cos\theta,t)}{\partial r}\ .
\label{Eq:adv}
\end{equation}

The Hamiltonian, on the right-hand side of Eqs.~\ref{Eq:QKEs1} and \ref{Eq:QKEs2}, includes the vacuum and neutrino self-interaction terms:
\begin{equation}
 H=H_{\mathrm{vac}} + H_{\nu\nu}\ ,
\end{equation}
where
\begin{equation}
    H_{\mathrm{vac}} = \frac{\omega}{2} \begin{pmatrix}
    -\cos 2\vartheta_{\mathrm{V}} & \sin 2\vartheta_{\mathrm{V}} \\
    \sin 2\vartheta_{\mathrm{V}} & \cos 2\vartheta_{\mathrm{V}}
    \end{pmatrix}\ ,
\end{equation}
with $\vartheta_{\mathrm{V}}$ being the vacuum mixing angle and $\omega = \Delta m^2/2E$ the vacuum frequency, $\Delta m^2$ the squared mass difference, and $E$ the neutrino energy. The correspondent vacuum Hamiltonian for antineutrinos is $\bar{H}_{\mathrm{vac}} = - H_{\mathrm{vac}}$. 
The self-interaction Hamiltonian is 
\begin{equation}
    H_{\nu\nu} = ~\mu_0 \int_{-1}^{1} \left[\rho(\cos\theta^\prime) - \bar{\rho}(\cos\theta^\prime) \right] \times \left(1-\cos\theta\cos\theta^\prime\right) d\cos\theta^\prime\ ,
\end{equation}
where $\mu_0$ is the self-interaction strength. 
For simplicity, we neglect the matter term in the Hamiltonian and instead use an effective smaller mixing angle. 

The last term on the right-hand side of Eqs.~\ref{Eq:QKEs1} and \ref{Eq:QKEs2} is the collision term including emission, absorption, and direction-changing scatterings: $\mathcal{C}(r,\cos\theta) = \mathcal{C}_{\mathrm{emission}} + \mathcal{C}_{\mathrm{absorb}} + \mathcal{C}_{\mathrm{dir-ch}}$, with
\begin{eqnarray}
    \mathcal{C}_{\text{emission}}^{i}&=&\frac{1}{\lambda_{\text {emission}}^{i}(r)}\ ,\\
    \mathcal{C}_{\text{absorb}}^{i} &=&-\frac{1}{\lambda_{\text{absorb}}^{i}(r)} \rho_{i i}(\cos \theta)\ ,\\
    \mathcal{C}_{\text{dir-ch}}^{i} &=&-\frac{2}{\lambda_{\text{dir-ch}}^{i}(r)} \rho_{i i}(\cos \theta) + \int_{-1}^1 \frac{1}{\lambda_{\text{dir-ch}}^{i}(r)} \rho_{i i}\left(\cos \theta^{\prime}\right) d\cos \theta^{\prime}\ ;
\label{Eq:col_term}
\end{eqnarray}
here, $\lambda^i(r)$ is the mean free path that depends on the neutrino flavor as summarized in Table \ref{Tab:mfp}. 

\begin{table}[t]
 \caption{Mean free paths of emission, absorption, and direction changing processes for the different flavors as functions of $\xi(r) = \exp(15-r)$. The radial dependence of the collision term corresponds to the one adopted in Case C of Ref.~\cite{Shalgar:2022lvv}.}
 \centering
 \begin{tabular}{|l|l|l|l|}
 \hline & $\nu_e$ & $\bar{\nu}_{{e}}$ & $\nu_x,~\bar{\nu}_x$ \\
 \hline \hline $\lambda_{\text{emission}_{\phantom{l}}}^{i}$[km] & $1 / [50 \xi(r)]$ & $1 / [30 \xi(r)]$ & $1 / [10 \xi(r)]$ \\
 \hline $\lambda_{\text {absorb }_{\phantom{l}}}^{i}$[km] & $1 / [50 \xi(r)]$ & $1 / [25 \xi(r)]$ & $1 / [10 \xi(r)]$ \\
 \hline $\lambda_{\text {dir-ch }_{\phantom{l}}}^{i}$[km] & $1 / [50 \xi(r)]$ & $1 / [25 \xi(r)]$ & $1 / [12.5 \xi(r)]$ \\
 \hline
 \end{tabular} 
 \label{Tab:mfp}
\end{table}

\begin{figure*}
\centerline{Without periodic boundaries\hspace{2.5cm} With periodic boundaries}
\vspace{0.5cm}
\centering
\includegraphics[width=0.4\textwidth]{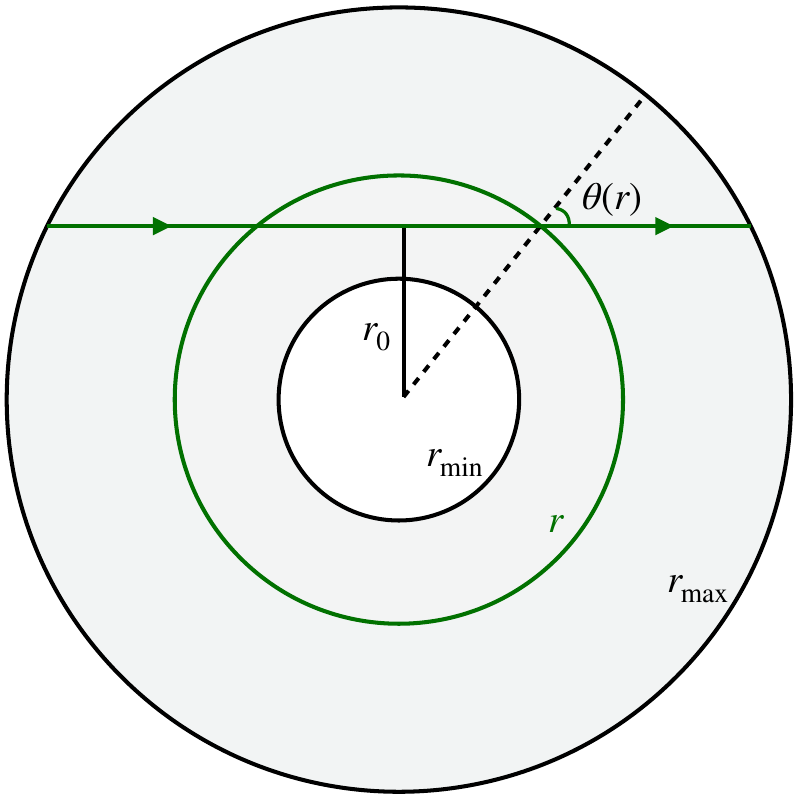} \hspace{0.7cm}
\includegraphics[width=0.4\textwidth]{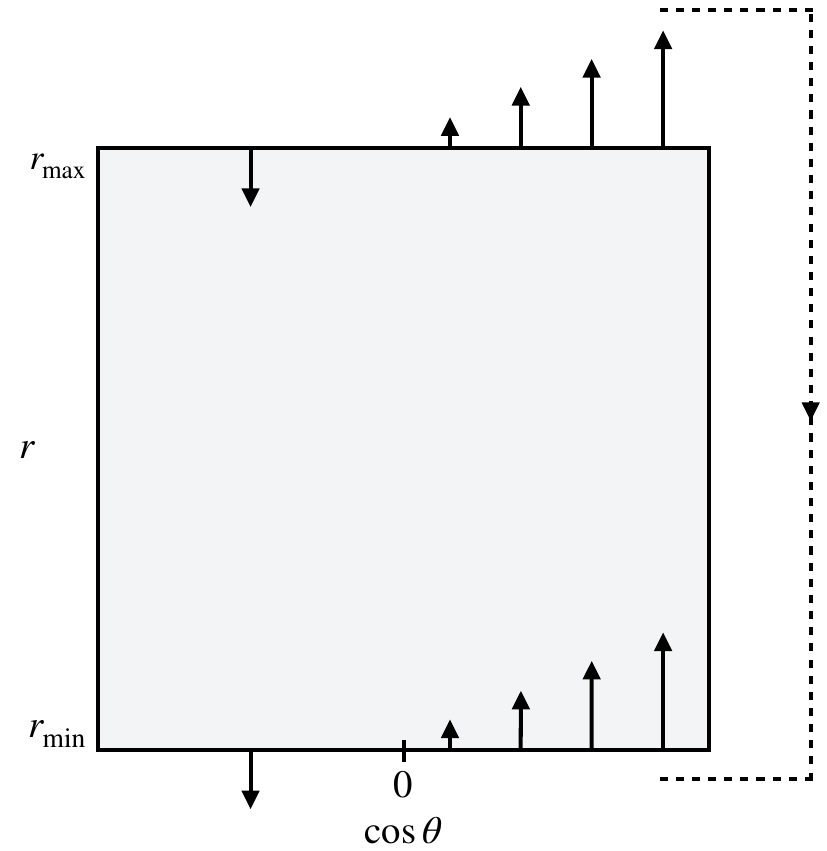}\\
\vspace{0.5cm}
\includegraphics[width=0.501\textwidth]{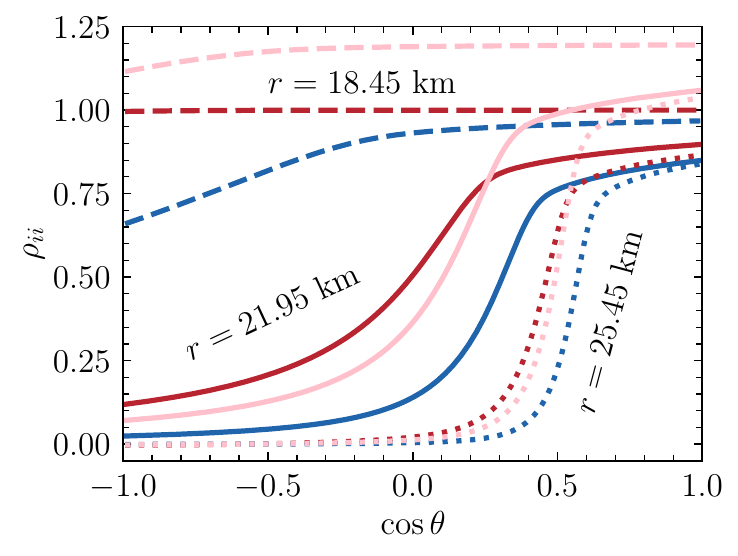}
\includegraphics[width=0.489\textwidth]{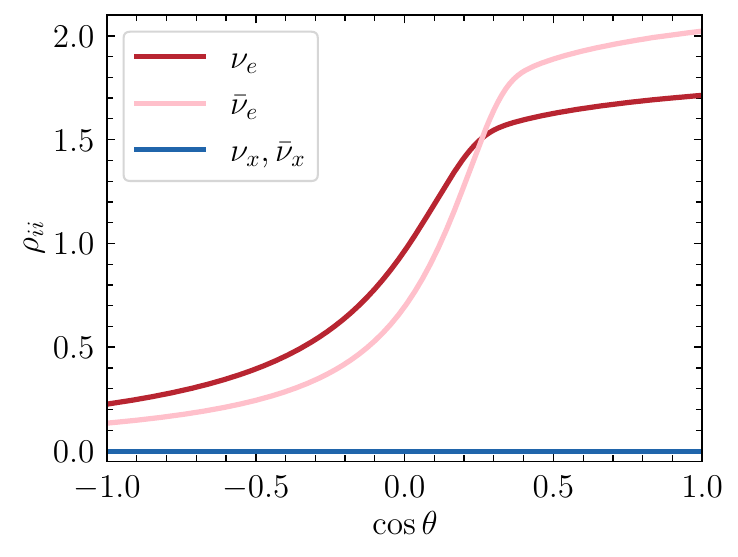}
\caption{{\it Top:} Sketch of the simulation shell without (on the left) and with (on the right) periodic boundary conditions. In the shell without periodic boundaries, the neutrino flavor evolution is investigated between $r_{\min}$ and $r_{\max}$. The neutrino angular distributions in the shell in the top left panel, at a given radius $r$, are the same throughout the shell (solid green circle). For any trajectory (thick solid green line), the angle with respect to the radial direction is $\theta(r)$. The radius $r_0$ marks the distance from the center of the simulation shell for a selected direction of propagation (see Fig.~\ref{Fig:no_perts}). In the absence of flavor conversion, the neutrino angular distributions become forward peaked as $r$ increases.
In the simulation box with periodic boundary conditions (top right panel), neutrinos can re-enter the box at $r_{\min}$ or $r_{\max}$; the direction of propagation of neutrinos is marked by the arrows. In this case, the neutrino angular distributions do not depend on $r$ in the absence of flavor conversion.
{\it Bottom:}
Corresponding angular distributions of $\nu_e$, $\bar\nu_e$, and $\nu_x$/$\bar\nu_x$ in the absence of flavor conversion. In the simulation shell without periodic boundaries, the angular distributions evolve as functions of radius becoming gradually forward peaked. In the simulation shell with periodic boundaries, the angular distributions do not evolve as functions of radius--we choose to use the distributions extracted from the simulation without periodic boundaries and without flavor conversion at $r = 21.95$~km as initial distributions throughout the periodic simulation box. }
\label{Fig:sim_sketch}
\end{figure*}

Our simulation shell is displayed in the top left panel of Fig.~\ref{Fig:sim_sketch} and is the same as the one used in Refs.~\cite{Shalgar:2022lvv,Shalgar:2022rjj}. It extends from $r_{\rm min} = 15$~km to $r_{\max}=30$~km, with $\cos{\theta} \in [-1,1]$. In the numerical solution of Eqs.~\ref{Eq:QKEs1} and \ref{Eq:QKEs2}, we rely on uniform grids with $150$ bins in $r$ and $\cos\theta$ which allow to achieve numerical convergence~\cite{Shalgar:2022lvv,Shalgar:2022rjj}. 
As for the other parameters entering Eqs.~\ref{Eq:QKEs1} and \ref{Eq:QKEs2}, we adopt $\mu_0 = 10^4~\mathrm{km}^{-1}$, $E = 20$~MeV, $\Delta m^2 = 2.5 \times 10^{-3}$~eV$^2$, and $\vartheta_{\mathrm{V}} = 10^{-3}$. Note that $\mu_0$ is the self-interaction strength at $r_{\min}$; the effective self-interaction strength decreases as $r$ increases because of the radial dependence of the neutrino number density and the average angle between the neutrinos.

We consider the simulation shell shown in the top left panel of Fig.~\ref{Fig:sim_sketch} to be initially empty and populate it with neutrinos and antineutrinos solving Eqs.~\ref{Eq:QKEs1} and \ref{Eq:QKEs2} with $H=\bar{H}=0$ 
(i.e.~without flavor conversion). The corresponding angular distributions obtained through this approach are shown in the left bottom panel of Fig.~\ref{Fig:sim_sketch} for representative $r$. One can see that, at small radii, the neutrino field is isotropic. At larger radii, as the neutrino density decreases, the angular distributions slowly become forward peaked. Such a trend holds for all flavors. Since the different neutrino flavors interact with matter at a different rate (see Table~\ref{Tab:mfp}), each flavor decouples from the medium at different radii. This behavior leads to the formation of electron lepton number (ELN) crossings.

We estimate that neutrino decoupling occurs where the flux factor is~\cite{Tamborra:2017ubu,Wu:2017drk}
\begin{equation}
 \mathcal{F}_{i} = \frac{\int_{-1}^{1}\rho_{ii}(r,\cos\theta)\cos\theta d\cos\theta}{\int_{-1}^{1}\rho_{ii}(r,\cos\theta) d\cos\theta} \approx \frac{1}{3}\ .
\end{equation}
Within the coupled region, the angular distribution of (anti)neutrinos is isotropic, implying $\mathcal{F}_{i} \rightarrow 0$. Conversely, in the free streaming regime, (anti)neutrinos exhibit a forward-peaked distribution, resulting in $\mathcal{F}_{i} \rightarrow 1$.

This classical steady-state solution is then used as the initial condition to solve Eqs.~\ref{Eq:QKEs1} and \ref{Eq:QKEs2} and investigate the flavor conversion physics until a quasi-steady state is reached at $t=50~\mu$s; the latter corresponds to the time that neutrinos take to cross the simulation shell traveling at the speed of light. We start the simulation with flavor conversion relying on  the classical steady-state solution to ensure numerical stability; however, the quasi-steady state configurations are independent of the details of the initial state used in the simulation.
As for the implementation of the boundary conditions, we follow Ref.~\cite{Shalgar:2022lvv}. 
At $r = r_{\min}$, the boundary condition is fixed by requiring a classical steady-state solution. At $r = r_{\max}$, the boundary condition depends on the direction of propagation of the neutrinos: for $\cos\theta > 0$ the simulation region determines the boundary, while we use a vanishing boundary condition for $\cos\theta \leq 0$.

\subsection{Simulation shell with periodic boundaries} 
Simulation shells with periodic boundary conditions are often employed in the literature, see e.g.~Refs.~\cite{Grohs:2022fyq,Richers:2021xtf,Richers:2022bkd,Bhattacharyya:2020jpj,Bhattacharyya:2022eed,Xiong:2023vcm,Wu:2021uvt,Martin:2019gxb,Martin:2021xyl,Abbar:2021lmm,Duan:2021woc}. 
This choice is motivated by the fact that, in the core of an astrophysical source, sufficiently small neighbor slices of matter are expected to have very similar properties; hence, one could gain computational time by focusing on the same shell and imposing periodic boundaries. However, the quasi-steady state solution depends on the choice of the boundary conditions~\cite{Nagakura:2022qko,Nagakura:2022kic,Nagakura:2022xwe,Shalgar:2022lvv,Shalgar:2022rjj,Zaizen:2023ihz,Sigl:2021tmj}. 

We consider the simulation shell with periodic boundaries displayed in the top right panel of Fig.~\ref{Fig:sim_sketch}; neutrinos moving in the forward direction ($\cos\theta > 0$) reach the edge of the simulation regime at $r_{\rm max}$ and reappear at $r_{\rm min}$ (and vice versa for neutrinos moving in the backward direction). 
When periodic boundary conditions are enforced, the neutrino number density can no longer decrease with radius, in order to avoid an abrupt change for neutrinos re-entering the simulation box at $r_{\rm max}$ (or $r_{\rm min}$). The periodic system is hence homogeneous, with the neutrino density being the same for all radii. Similarly, the polar angle $\theta$ describing the propagation direction of neutrinos does not depend on radius. This implies that the first term is zero in Eq.~\ref{Eq:adv}. 

 For the periodic case, we consider $\mathcal{C}= \bar{\mathcal{C}} = 0$ in Eqs.~\ref{Eq:QKEs1} and \ref{Eq:QKEs2} and assume as angular distributions the ones taken from the classical solution of the simulation shell without periodic boundaries (bottom left panel of Fig.~\ref{Fig:sim_sketch}) extracted at 
$r = 21.95$~km, when neutrinos have decoupled from matter. We only use the angular distributions for $\nu_e$ and $\bar\nu_e$ and set the distributions for $\nu_x$ and $\bar\nu_x$ to zero for simplicity.
The bottom right panel of Fig.~\ref{Fig:sim_sketch} shows the angular distributions, normalized such that $\int_{-1}^{1} \rho_{ee} d\cos\theta / \int d\cos\theta = 1$. 
In order to investigate the flavor evolution, we solve Eqs.~\ref{Eq:QKEs1} and \ref{Eq:QKEs2} using $400$ radial bins and $150$ angular bins and evolve the system until $50~\mu$s. All other input parameters are chosen to be the same as the ones in the simulation box with periodic boundaries.

\section{Neutrino flavor evolution without perturbations}
\label{Sec:no_perts}
In this section, we investigate the quasi-steady state configurations of neutrino flavor obtained for the simulation shells with and without periodic boundaries (see Fig.~\ref{Fig:sim_sketch}) without introducing any perturbation in the neutrino field.
\subsection{Simulation without periodic boundaries}
\begin{figure}
\centerline{No periodic boundaries, no perturbations}
\vspace{0.5cm}
    \includegraphics[width=0.99\textwidth]{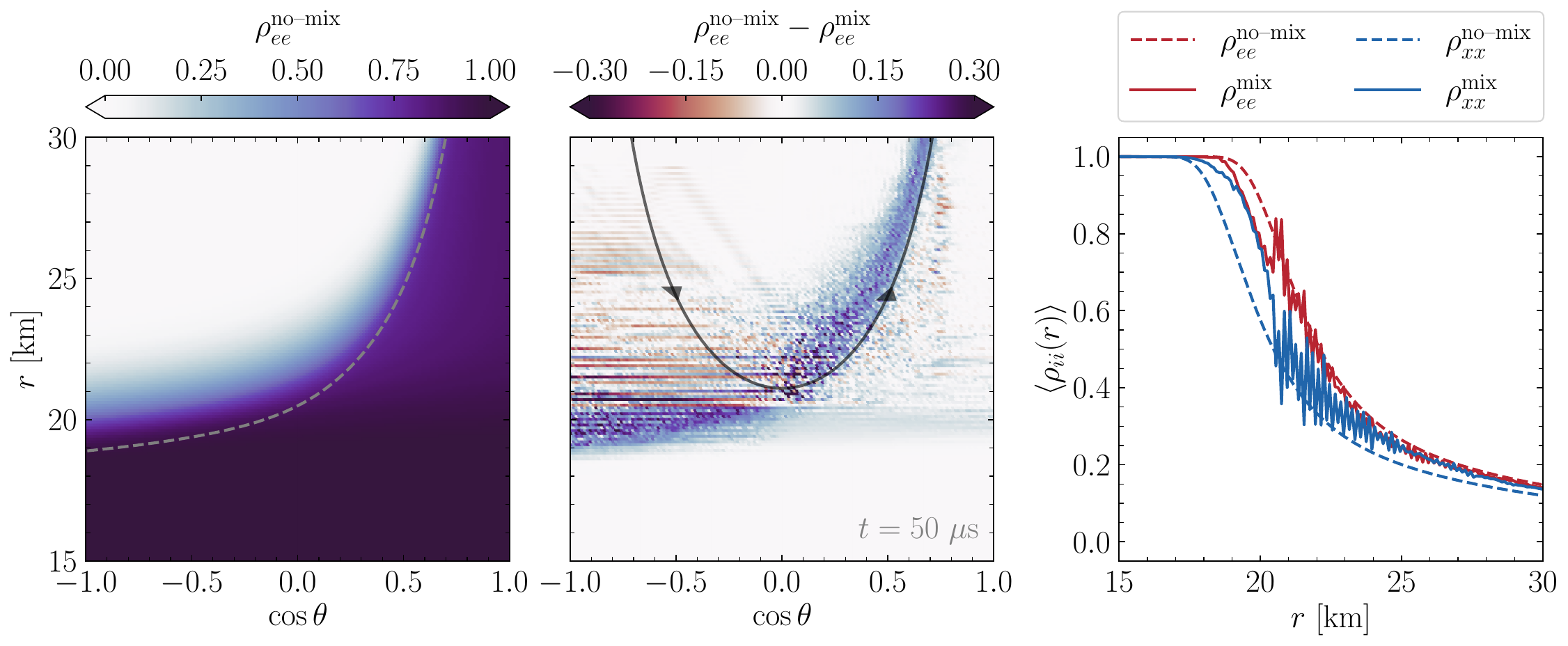}
    \caption{Quasi-steady state neutrino flavor configuration without perturbations and for the simulation shell without periodic boundaries at $50~\mu$s. {\it Left:} Steady-state neutrino flavor configuration in the absence of flavor mixing (obtained solving Eqs.~\ref{Eq:QKEs1} and \ref{Eq:QKEs2} with $H=\bar{H}=0$). The light gray dashed line highlights the locus of ELN crossings.
    {\it Middle:} Contour plot of the difference of $\rho_{ee}$ with and without flavor conversion. The black line represents the trajectory followed by a neutrino emitted at $r = 21.95$ km and $\cos\theta=0.27$. Flavor conversion develops in the proximity of the ELN crossings.
    {\it Right:} Angle-averaged radial profile of $\rho_{ee}$ (in red) and $\rho_{xx}$ (in blue). The dashed (solid) lines indicate angle-averaged number densities obtained in the absence (in the presence) of flavor conversion. Flavor equipartition is not achieved.}
 \label{Fig:no_perts}
\end{figure}

The left panel of Fig.~\ref{Fig:no_perts} shows the steady-state configuration of $\rho_{ee}$ obtained for the non-periodic shell (left panel of Fig.~\ref{Fig:sim_sketch}). 
To investigate the impact of flavor conversion, Eqs.~\ref{Eq:QKEs1} and \ref{Eq:QKEs2} are solved until 
$50~\mu$s when a quasi-steady state is reached. The middle panel of Fig.~\ref{Fig:no_perts} shows a contour plot of the difference of $\rho_{ee}$ between the cases when neutrino flavor conversions are taken into account (mix) and with no flavor conversions (no–mix) extracted at $50~\mu$s. 
The blue region in the middle panel of Fig.~\ref{Fig:no_perts} presents an overall deficit of electron neutrinos with respect to the case without flavor conversion. This is because flavor conversion takes place triggered by ELN crossings (see white dashed line in the left panel of Fig.~\ref{Fig:no_perts}).
Neutrinos that undergo flavor transformation at one radius are then transported at larger radii due to advection. Hence, the region affected by flavor conversion expands as a function of time due to advection and collisions. 

Following the procedure outlined in Ref.~\cite{Duan:2006an}, the trajectory of a free-streaming neutrino in the simulation shell is given by:
\begin{equation}
    r(\cos\theta) = \frac{\sin\theta_0}{\sin\theta}r_{\nu}\ ,
\label{Eq:trajectory}
\end{equation}
where $\theta_0$ and $r_{\nu}$ are the emission angle and radius, respectively. If neutrinos stream freely, without experiencing any interactions, the minimum radius that they can reach in the simulation shell following Eq.~\ref{Eq:trajectory} is $r_0$, as depicted in the left panel of Fig.~\ref{Fig:sim_sketch}.
The trajectory of a neutrino emitted at $r=21.95$~km and $\cos\theta=0.27$ where the ELN crossing takes place is plotted on top of the contour plot in Fig.~\ref{Fig:no_perts} to guide the eye.

The right panel of Fig.~\ref{Fig:no_perts} shows the radial profile of the angle-averaged $\rho_{ee}$ and $\rho_{xx}$, defined as 
$\langle \rho_{ii}(r)\rangle = {\int_{-1}^{1} \rho_{ii}(r,\cos\theta)d\cos\theta}/{\int_{-1}^{1} d\cos\theta}$.
One can see that flavor conversion affects $\langle \rho_{ee}\rangle$ and $\langle \rho_{xx}\rangle$ across radii. Because of flavor conversion, $\langle\rho_{ee}\rangle$ and $\langle\rho_{xx}\rangle$ deviate from their correspondent ones obtained in the absence of flavor conversion (cf.~solid vs.~dashed lines in the plot); however flavor equipartition is not achieved.

\subsection{Simulation with periodic boundaries}
\begin{figure}
\centerline{Periodic boundaries, no perturbations}
\vspace{0.5cm}
 \includegraphics[width=0.99\textwidth]{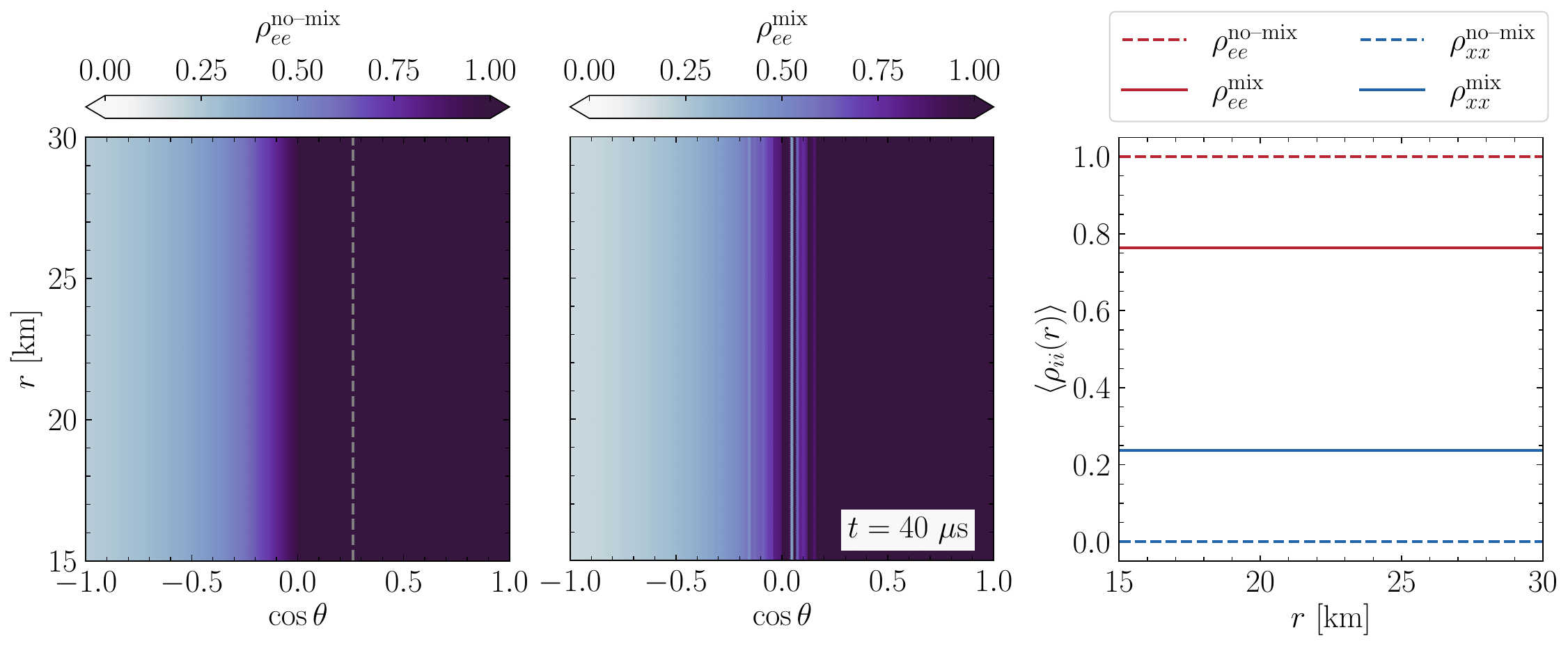}
 \caption{Quasi-steady state neutrino flavor configuration without perturbations and for the simulation shell with periodic boundaries. \textit{Left}: Steady-state neutrino flavor configuration in the absence of flavor mixing. The light gray dashed line represents the ELN crossing loci.  \textit{Middle}: Contour plot of $\rho_{ee}$ in the presence of flavor mixing at $40~\mu$s. The ELN crossings lead to the development of flavor conversion in their surroundings. \textit{Right}: Angle-averaged radial profiles of $\rho_{ee}$ and $\rho_{xx}$ obtained in the absence (dashed lines) and in the presence (solid lines) of flavor conversion. As time increases, in the presence of flavor conversion, the angle-averaged $\rho_{ee}$ and $\rho_{xx}$ tend to approach each other due to flavor conversion.}
 \label{Fig:no_perts_periodic}
\end{figure}

Figure~\ref{Fig:no_perts_periodic} shows the flavor outcome in the case of the simulation shell with periodic boundaries. The left panel represents the initial flavor configuration based on the angular distributions extracted from the classical steady-state solution without periodic boundaries (see Fig.~\ref{Fig:sim_sketch}).
The middle panel shows the neutrino flavor configuration at $t=40~\mu$s and in the presence of flavor conversion. One can see that flavor conversion occurs in the proximity of the ELN crossing loci. 
As time increases flavor conversion spreads through all angular modes.
The right panel shows the corresponding angle-averaged radial profile of each neutrino flavor. Initially, $\langle \rho_{ee}(r)\rangle = 1$ and $\langle \rho_{xx}(r)\rangle = 0$, while $\langle \rho_{ee}(r)\rangle$ and $\langle \rho_{xx}(r)\rangle$ approach each other due to flavor conversion (solid lines) at $t=40~\mu$s. Note that there is no radial dependence since the system is homogeneous.

\section{Neutrino flavor evolution with perturbations}
\label{Sec:impact_of_perts}
In this section, we first introduce the approach adopted to perturb the neutrino field. Then, we investigate the quasi-steady state configurations of neutrino flavor obtained for the simulation shells with and without periodic boundaries (see Fig.~\ref{Fig:sim_sketch}) in the presence of Gaussian perturbations. 

\subsection{Perturbations of the neutrino field}
\label{Subsec:perts_def}
In order to mimic anisotropies in the neutrino background, the density matrix is perturbed as follows 
\begin{equation}
 \rho = \begin{pmatrix}
 \rho_{ee} (1+\xi P_{ee}) & \rho_{ex} + \frac{1}{2}\left(\rho_{ee} + \rho_{xx}\right)\xi P_{ex} \\
 \rho_{xe} + \frac{1}{2}\left(\rho_{ee} + \rho_{xx}\right)\xi P_{xe} & \rho_{xx} (1+\xi P_{xx})
 \end{pmatrix}\ ,
\end{equation}
where $P_{ij} \equiv P_{ij}(r,\cos\theta, t)$ and $\xi$ is the strength of the perturbations. 

At the time $t$, the Gaussian perturbation $P$ of the neutrino field in $\left(r_k,\cos{\theta}_l\right)$ is
\begin{equation}
 P_{ij}(r_k,\cos{\theta}_l, t) = \frac{1}{N}~\exp\left(-\frac{(r_k-r_0)^2}{2 \sigma_r^{ij 2}(r_0)} - \frac{(\cos{\theta}_l-\cos{\theta}_0)^2}{2 \sigma_{\cos{\theta}}^2(\cos\theta_0)}\right) \delta(t-t^\star)\ ,
\label{Eq:spat_perts}
\end{equation}
with the normalization factor $N$ being such that $\int P_{ij}(r_k,\cos{\theta}_l) d\cos{\theta_l} dr_k = 1$ and $\left(r_0,\cos{\theta}_0\right)$ representing the center of the perturbation. The angular and radial widths of the Gaussian are given by $\sigma_{\cos{\theta}}$ and $\sigma_r^{ij} $, respectively. The former is defined as $\sigma_{\cos{\theta}} = f_{\cos\theta} (2 + \cos{\theta})$, where $f_{\cos\theta}$ determines the angular range where the perturbation is applied. The diagonal component of the radial width depends on the mean free path:
\begin{equation}
 \sigma_r^{ii} = f_r \lambda^{i}(r)\ ,
\end{equation}
and the off-diagonal component is defined as
\begin{equation}
 \sigma_r^{ij} = f_r\left(\frac{1}{2\lambda^{i}(r)} + \frac{1}{2\lambda^{j}(r)}\right)^{-1}\ .
\end{equation}
Here $1/\lambda^{i}(r) = 1/\lambda^{i}_{\text{emission}}(r) + 1/\lambda^{i}_{\text{absorb}}(r) + 1/\lambda^{i}_{\text{dir-ch}}(r)$; similar definitions hold for antineutrinos. For what concerns the radial extension of the perturbation, a scaling factor $f_r$ is adopted. The delta function in Eq.~\ref{Eq:spat_perts} indicates that each perturbation of the neutrino field is injected at a specific time step $t^\star$, randomly chosen from a uniform distribution, such that perturbations appear three times at the same location over the simulation duration. 
To make sure that the perturbations are not injected too close to the end of the simulation, when a quasi-steady state configuration is reached, we set an upper limit on the perturbation injection time at $t\simeq 40~\mu$s. 
This ensures that, even if the perturbation did not have enough time to move out of the box, the impact of all perturbations on flavor conversion can be explored.

\begin{figure*}
\centerline{No periodic boundaries, with perturbations}
 \vspace{0.5cm}
\centerline{\hspace{-1.7cm}Case N1 \hspace{4.6cm} Case N2}
 \includegraphics[width=0.99\textwidth]{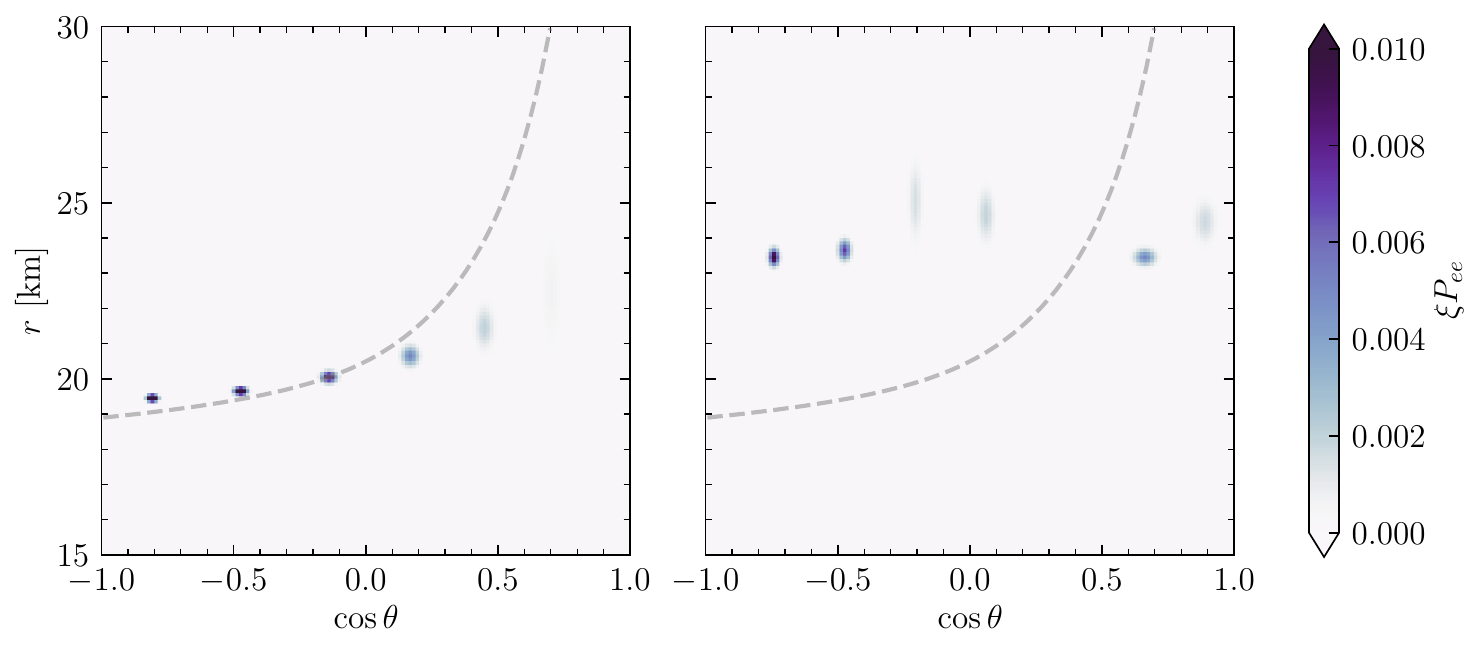}
\caption{Contour plot of the perturbation $\xi P_{ee}$ (for $\xi=0.1$) in the plane spanned by $\cos\theta$ and $r$ for Cases N1 (left panel) and N2 (right panel). The spatial perturbations are located inside and outside the ELN crossing region for Cases N1 and N2, respectively. Each perturbation appears three times at the same location over the simulation duration at a random time step (see main text for details). The gray line shows the ELN crossing locus.
The locations of the perturbations are the same for the simulations with and without periodic boundaries for simplicity, although the ELN crossing loci are different. }
\label{Fig:pert_exAB}
\end{figure*}

Since the location and strength of the perturbations may affect the development of flavor conversion, we consider two different cases. In Case N1, we place six perturbations in the proximity of the location of the ELN crossings in our simulation shell, as displayed in the left panel of Fig.~\ref{Fig:pert_exAB}, where all perturbations $\xi P_{ee}$ (with $\xi=0.1$) are shown side by side to illustrate their location with respect to the ELN crossing region (a similar trend occurs for the other neutrino flavors, not shown here). 
In Case N2, the six perturbations are placed at random radial and angular locations in the simulation shell, outside the region of ELN crossings, but above the radius of neutrino decoupling (see right panel of Fig.~\ref{Fig:pert_exAB}). In fact, when neutrinos are in the trapping regime, any perturbation quickly dissipates and is diluted, with no impact on the flavor evolution. The values of $\xi$, $f_r$, and $f_{\cos\theta}$ are summarized in Table~\ref{Tab:pert_params} for Cases N1 and N2. For the simulation with periodic boundaries, the ELN crossing locus is independent of radius since the initial distribution is homogeneous. However, the perturbations are injected in the same locations as shown in Fig.~\ref{Fig:pert_exAB} in order to compare the impact of the different types of boundary conditions. 
\begin{table}[t]
\caption{Spatial perturbations of the neutrino field (see Eq.~\ref{Eq:spat_perts}).}
\centering
\begin{tabular}{|l|c|c|}
\hline
& {Case N1} & {Case N2} \\ \hline \hline
$\xi$ & $0.1$ & $0.1$ 
\\ \hline
$f_r$ & $0.1$ & $0.005$ 
\\ \hline
$f_{\cos\theta}$ & $0.01$ & $0.01$ 
\\ \hline
\end{tabular}
\label{Tab:pert_params}
\end{table}

\subsection{Simulation shell without periodic boundaries}

\begin{figure*}
\centerline{No periodic boundaries, with perturbations}
 \vspace{0.5cm}
\centerline{\hspace{-1.22cm}Case N1 \hspace{4.2cm} Case N2}
 \includegraphics[width=0.99\textwidth]{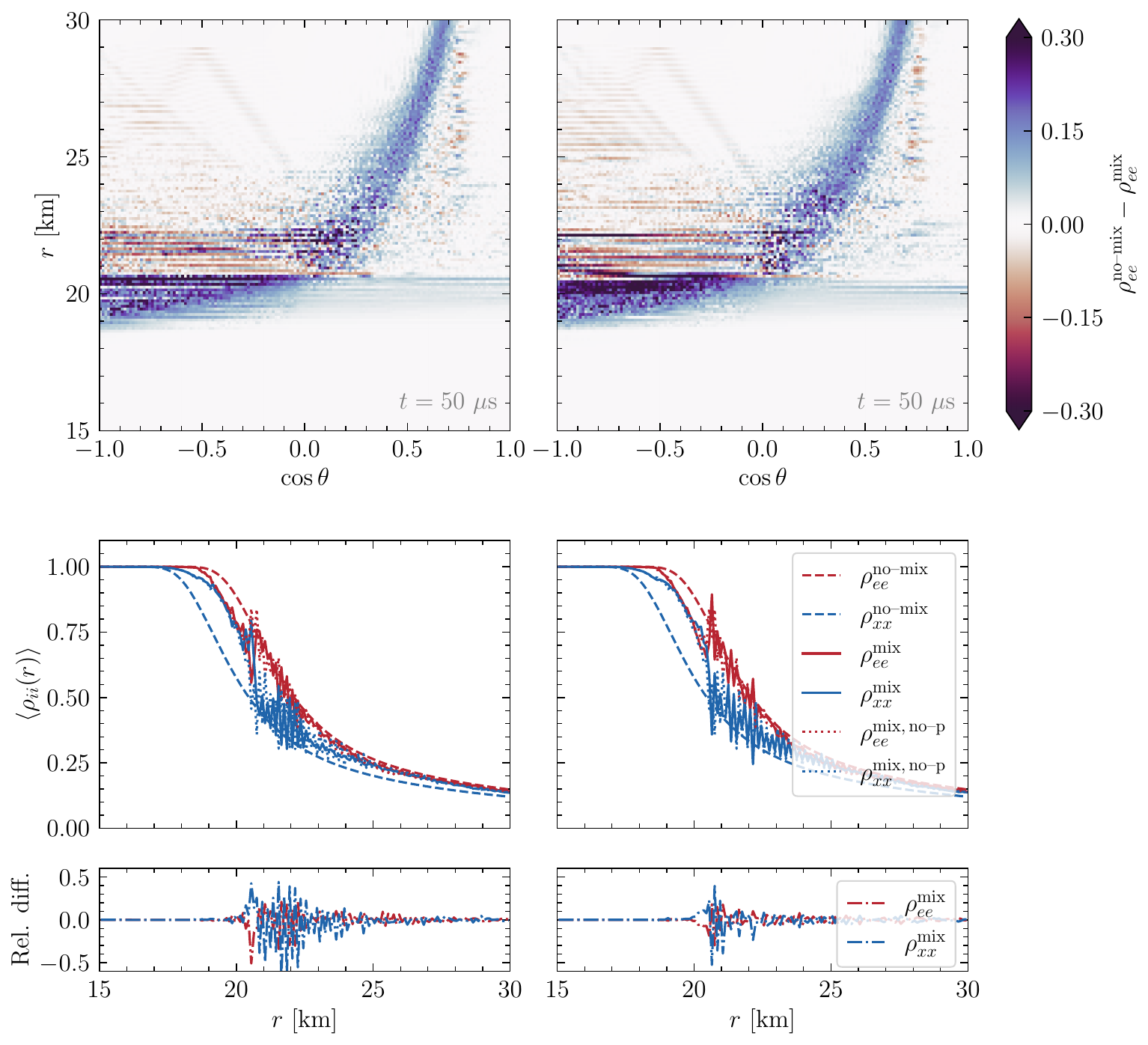}
\caption{Quasi-steady state neutrino flavor configuration with perturbations and no periodic boundary conditions for Cases N1 (left panels) and N2 (right panels) at $50~\mu$s. {\it Top panels:} Contour plots of the difference between $\rho_{ee}$ with and without flavor conversion in the plane spanned by $\cos\theta$ and $r$. Flavor conversion mostly develops in the surroundings of the ELN crossings. The red spots represent regions underdense in $\nu_e$ due to flavor conversion with respect to the case without flavor conversion. {\it Bottom panels:} Radial evolution of the angle-averaged $\rho_{ee}$ (in red) and $\rho_{xx}$ (in blue) for the cases without flavor conversion (dashed lines), with flavor conversion and perturbations (solid lines), as well as with flavor conversion and without perturbations (dotted lines). The relative difference between the solid and dotted lines is shown in the bottom panels.}
\label{Fig:pert_AB}
\end{figure*}

The top panels of Fig.~\ref{Fig:pert_AB} show contour plots of $\rho_{ee}$ for Cases N1 and N2 comparing the number density with and without flavor conversion at $t=50~\mu$s (animations of the flavor evolution are available \href{https://sid.erda.dk/share_redirect/CjZUbh90L0/index.html}{here}). As for the case without perturbations shown in Fig.~\ref{Fig:no_perts}, the neutrino field is isotropic at small radii and becomes forward peaked as neutrinos decouple from the medium.
The red spots in the contour plots represent regions with an underdensity of $\nu_e$ due to flavor conversion and may have an unphysical origin; however, we have checked that they do not affect the overall flavor evolution.

The middle panels of Fig.~\ref{Fig:pert_AB} show the quasi-steady state, angle-averaged neutrino number densities for the different flavors as functions of radius. We compare the solution without flavor conversion (dashed lines), the one with flavor conversion and perturbations (solid lines), and the one with flavor conversion but without perturbations (dotted lines).
It can be seen that for both cases with and without perturbations for Cases N1 and N2, no flavor equipartition is achieved between the $e$ and $x$ flavors. 

The bottom panels of Fig.~\ref{Fig:pert_AB} display the relative difference between the dashed and dotted lines, defined as $(\rho_{ii}^{\mathrm{mix}}-\rho_{ii}^{\mathrm{mix,\, no-p}})/\rho_{ii}^{\mathrm{mix}}$. We see that the relative difference among the flavor distributions of the simulations with and without perturbations, in the presence of flavor conversion, can be less than $10\%$ on average. 
We have also tested Cases N1 and N2 with $\xi=10^{-3}$ and $\xi=10^{-5}$ and found smaller differences with respect to the case without perturbations (results not shown here).
Hence, while some specific configurations of the angular distributions for the different flavors could result in flavor equipartition (e.g., see the one adopted in Ref.~\cite{Shalgar:2022rjj}), the latter is not a general outcome. Perturbations of the neutrino field can induce changes in the flavor conversion physics, but they are not responsible for pushing the system versus flavor equipartition, confirming the preliminary results of Refs.~\cite{Shalgar:2022rjj,Shalgar:2022lvv,Padilla-Gay:2022wck,Nagakura:2022qko,Nagakura:2022kic,Nagakura:2022xwe}.

\begin{figure*}
\centerline{No periodic boundaries, with perturbations, Case N2}
\vspace{0.5cm}
\includegraphics[width=0.99\textwidth]{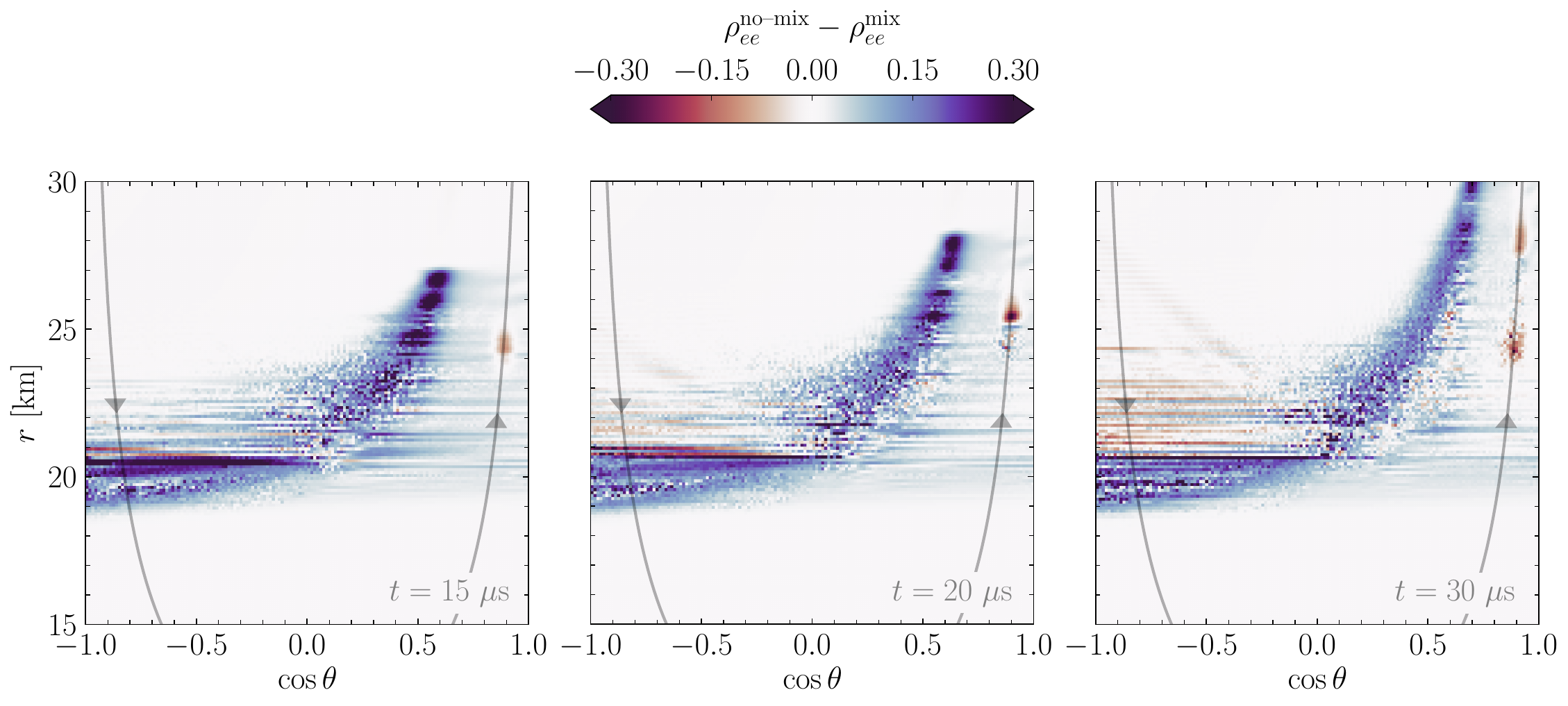}
\caption{Same as the top right panel of Fig.~\ref{Fig:pert_AB}, but for selected snapshots ($t=15$, $20$, and $30$~$\mu$s) to highlight how the perturbations of the neutrino field stream out of the simulation shell. The rightmost perturbation from Case N2 (see top right panel of Fig.~\ref{Fig:pert_exAB}) is increased to $\xi=10$ for visibility, and one can see how it moves along the neutrino trajectory (Eq.~\ref{Eq:trajectory}).
In the first time snapshot (left panel, $t = 15~\mu$s), a perturbation of the neutrino field is placed at $r\simeq 24.5$~km and $\cos\theta \simeq 0.9$, away from the region of ELN crossings. In the middle panel, at $t=20~\mu$s, the perturbation has moved to a larger radius ($r\simeq 26$~km). In the right panel, as this perturbation approaches the outer edge of the simulation shell, a new perturbation appears at $r\simeq 24.5$~km and $\cos\theta \simeq 0.9$. The perturbation moves out of the simulation shell before diffusing and affecting flavor conversion (see \href{https://sid.erda.dk/share_redirect/CjZUbh90L0/index.html}{animation}).}
\label{Fig:pert_ill}
\end{figure*}

In order to highlight the interplay among perturbations, advection, and flavor conversion, we focus on Case N2 and show snapshots of the flavor evolution in 
Fig.~\ref{Fig:pert_ill}. 
The perturbation is here considered for $\xi=1$ such that it is well visible on the contour plot with respect to the ELN crossing region.
At $t = 15~\mu$s, a perturbation is placed at $r \simeq 25$~km and $\cos\theta \simeq 0.9$ (left panel). Such perturbation tends to move upwards to larger radii as time increases because of neutrino advection. The expected trajectory of the perturbation is represented by the gray line (see Eq.~\ref{Eq:trajectory}). Since the perturbation is placed at $\cos\theta > 0$ it tends to reach  larger radii with time.
The middle panel of Fig.~\ref{Fig:pert_ill} displays a time snapshot at $t = 20~\mu$s where the perturbation is located at $r \simeq 26$~km. 
In the right panel of Fig.~\ref{Fig:pert_ill}, the perturbation has moved to $r \simeq 28$~km at $t = 30~\mu$s, while a second perturbation is introduced at $r \simeq 25$~km where the first one was originally placed. The perturbation of the neutrino field does thus not affect the development of flavor conversion when placed outside of the ELN crossing region. These perturbations move forward, because of neutrino advection, faster than they could diffuse, affecting the flavor evolution.

Note that we have also repeated the simulations used to produce Fig.~\ref{Fig:pert_AB} for Cases N1 and N2, introducing perturbations only in the off-diagonal elements of the density matrix such that
\begin{equation}
 \rho = \begin{pmatrix}
 \rho_{ee} & \rho_{ex} + \frac{1}{2}\left(\rho_{ee} + \rho_{xx}\right)\xi P_{ex} \\
 \rho_{xe} + \frac{1}{2}\left(\rho_{ee} + \rho_{xx}\right)\xi P_{xe} & \rho_{xx}
 \end{pmatrix}\ ,
\end{equation}
with $\xi P_{ij}$ defined as in Sec.~\ref{Subsec:perts_def}.
We obtain quasi-steady state configurations (results not shown here) which do not show flavor equipartition. This confirms that, in the absence of periodic boundaries, the presence of perturbations in the neutrino field does not guarantee the achievement of flavor equipartition.

\subsection{Simulation shell with periodic boundaries}
\begin{figure*}
\centerline{Periodic boundaries, with perturbations, Case N1}
\vspace{0.5cm}
\includegraphics[width=0.99\textwidth]{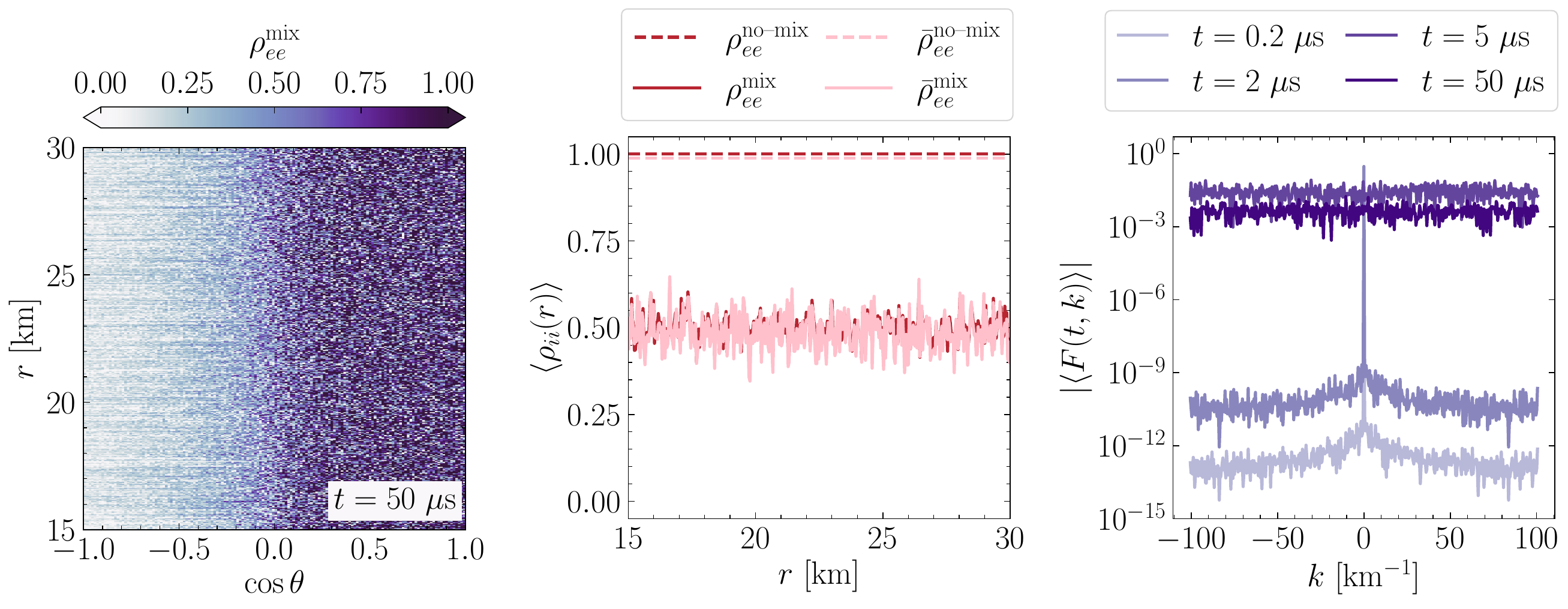}

\vspace{0.5cm}
\centerline{Periodic boundaries, with perturbations, Case N2}

\vspace{0.5cm}
\includegraphics[width=0.99\textwidth]{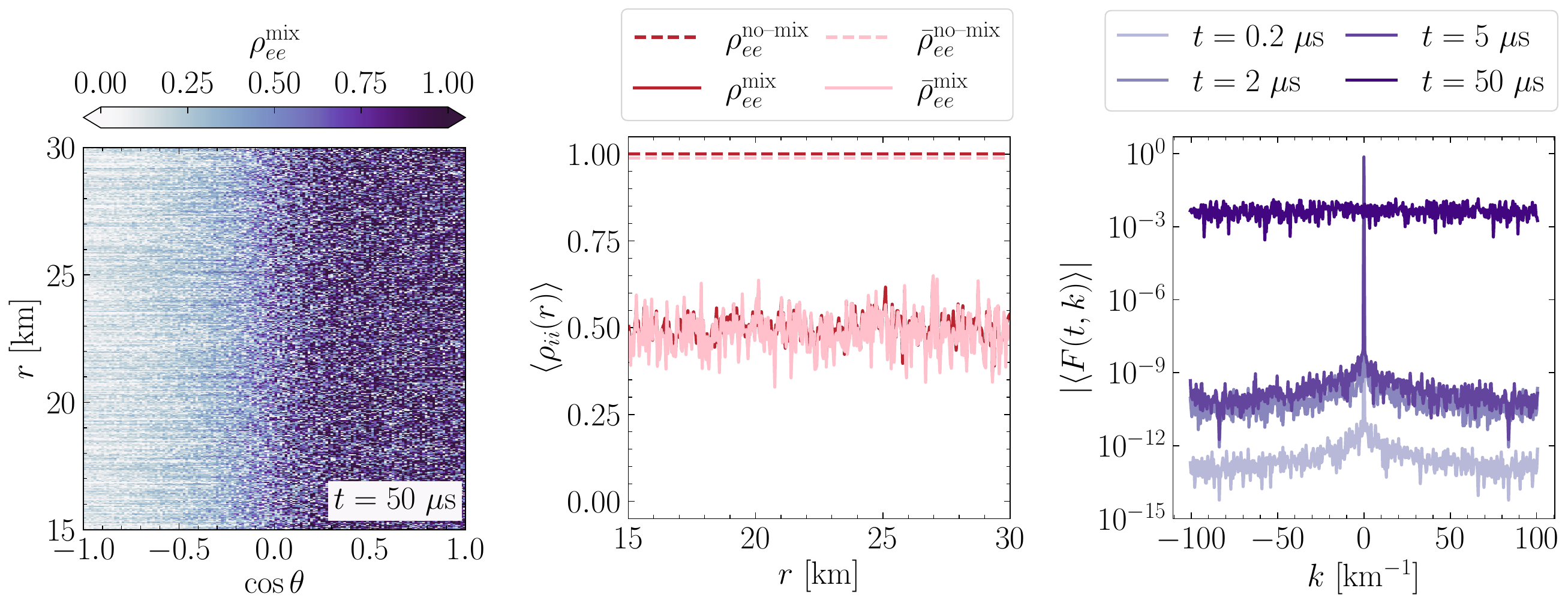}
\caption{Quasi-steady state neutrino configuration with perturbations and periodic boundary conditions, for Cases N1 (top panels) and N2 (bottom panels). {\it Left panels:} Contour plot of $\rho_{ee}$ in the presence of flavor conversion at $t=50~\mu$s in the plane spanned by $\cos\theta$ and $r$. 
{\it Middle panels:} Angle-averaged radial profile of $\rho_{ee}$ (red) and $\bar\rho_{ee}$ (pink) with (solid lines) and without (dashed lines) flavor conversion. 
{\it Right panels:} Angle-averaged Fourier power spectrum of the off-diagonal component of the density matrix $\rho_{ij}$ (see Eq.~\ref{eq:Fourier}) for different time snapshots. 
For both Cases N1 and N2, flavor conversion spreads across the simulation shell and flavor equipartition is achieved. Note that flavor conversion cascades down to smaller and smaller scales, until all Fourier modes are uniformly occupied.}
\label{Fig:pert_periodic_gaussian}
\end{figure*}

Figure~\ref{Fig:pert_periodic_gaussian} displays the quasi-steady state flavor configuration obtained for Cases N1 and N2 for the simulation shell with periodic boundary conditions and in the presence of perturbations. The left panels represent isocontours of the quasi-steady state $\rho_{ee}$ at $t=50~\mu$s (see \href{https://sid.erda.dk/share_redirect/CjZUbh90L0/index.html}{animation}). We see that $\rho_{ee}$ evolves developing small scale structures with large fluctuations between $0$ and $1$. 
Moreover, in the backward (forward) direction, $\rho_{ee}$ appears on average in light (dark) blue, meaning that there are fewer (more) electron neutrinos. This is due to the forward peakedness of the angular distribution of electron neutrinos. 
On average, flavor equipartition is achieved as visible in the middle panels of Fig.~\ref{Fig:pert_periodic_gaussian} where the angle-averaged $\rho_{ee}$ and $\bar\rho_{ee}$ fluctuate around $0.5$ as  functions of the radius.

In order to investigate the diffusion of flavor waves at small scales, we compute the direction-averaged Fourier spectrum, following Ref.~\cite{Richers:2022bkd}:
\begin{equation}
\label{eq:Fourier}
 \langle F(t,k) \rangle = \frac{1}{L} \int_{-1}^{1} \int_{0}^{L} \rho_{ij}(r,\cos\theta,t)e^{-ikr}dr~d\cos\theta\ ,
\end{equation}
where $L=r_{\max}-r_{\min}$ is the length of the simulation shell and $k=2\pi n / L$, with $n$ the order of the Fourier mode. 
The Fourier spectrum is shown in the right panels of Fig.~\ref{Fig:pert_periodic_gaussian} as a function of $k$ for different time snapshots.
As time increases, the flavor unstable modes grow until they saturate, leading to all Fourier modes to contribute increasingly. 
Comparing the top and bottom right panels of Fig.~\ref{Fig:pert_periodic_gaussian}, we see that the rate at which large $k$ are populated differs from Cases N1 and N2; however, at $t=50~\mu$s, the Fourier spectra are comparable; the reason for this difference is that the random time steps at which the perturbations are injected occur for Case N1 on average earlier than for Case N2, favoring the spread of structures at small scales. 

We can understand whether flavor equipartition is achieved in the simulation shell with periodic boundaries only for specific shapes of the initial angular distributions. We explore different configurations in Appendix \ref{App:other_ang_dist} and find that flavor equipartition is reached in the whole box, with angle-averaged radial profiles comparable to those of the benchmark angular distributions of the middle panel of Fig.~\ref{Fig:pert_periodic_gaussian}. However, the contour plots suggest that there is a difference between the two configurations across the angular range. This is explored in  Sec.~\ref{Sec:ang_dist_dependence}.

Our findings for the simulation shell with periodic boundary conditions are in good agreement with the ones of Ref.~\cite{Richers:2022bkd}, which employed periodic boundaries and perturbations. 
However, note that the Fourier spectrum in the right panels of Fig.~\ref{Fig:pert_periodic_gaussian} peaks at $k \simeq 0$, while this is not the case in Fig.~3 of Ref.~\cite{Richers:2022bkd}; this difference is due to the shape of the angular distributions. 
In fact, the position of the peak of the Fourier spectrum in $k$-space depends on  how forward peaked the angular distributions are. The distributions adopted in Ref.~\cite{Richers:2022bkd} are more  forward-peaked than the ones used in this work; this implies that flavor instabilities are more likely to  develop in the forward direction, with a related increase of the Fourier spectrum at negative $k$ (see  Fig.~3 of Ref.~\cite{Richers:2022bkd}). In our case, the angular distributions are less forward peaked. Therefore a peak in the Fourier spectrum is  expected around negative $k$ close to $k \simeq 0$; yet, the location of the initial peak does not affect the cascade of flavor waves to small scales. We also note that the Fourier spectrum displayed in Fig.~3 of Ref.~\cite{Richers:2022bkd} is not flat at later times. Running a simulation with identical setup to the one of Ref.~\cite{Richers:2022bkd}, we conclude that this trend is due to the fact that their simulation has not been run long enough for the perturbations to evolve across all length scales (evolving our simulation up to $t=50000 \mu^{-1}$, we do observe a flattening of the power spectrum; results not shown here).
It should be noted that Fig.~3 of Ref.~\cite{Richers:2022bkd} has Fourier modes in units of $\mu^{-1}$ and a larger value of $L$, which is why the range in $k$ is smaller than the one adopted in this paper. Yet, the two figures are comparable, when calculating the order of the Fourier modes.

Instead of the Gaussian perturbations, we also investigate the impact on the flavor outcome of a cosine fluctuation along the radial direction, as discussed in Appendix \ref{App:radial_perts}. We find a similar trend with respect to the one illustrated above and conclude that the occurrence of flavor equipartition does not depend on the specific nature of the perturbations.

\subsection{Simulation shell with periodic boundaries: Dependence of the quasi-steady state flavor configuration on the initial angular distribution}
\label{Sec:ang_dist_dependence}
\begin{figure*}
\centerline{Periodic boundaries, with perturbations, Case N1}
\vspace{0.5cm}
 \includegraphics[width=0.99\textwidth]{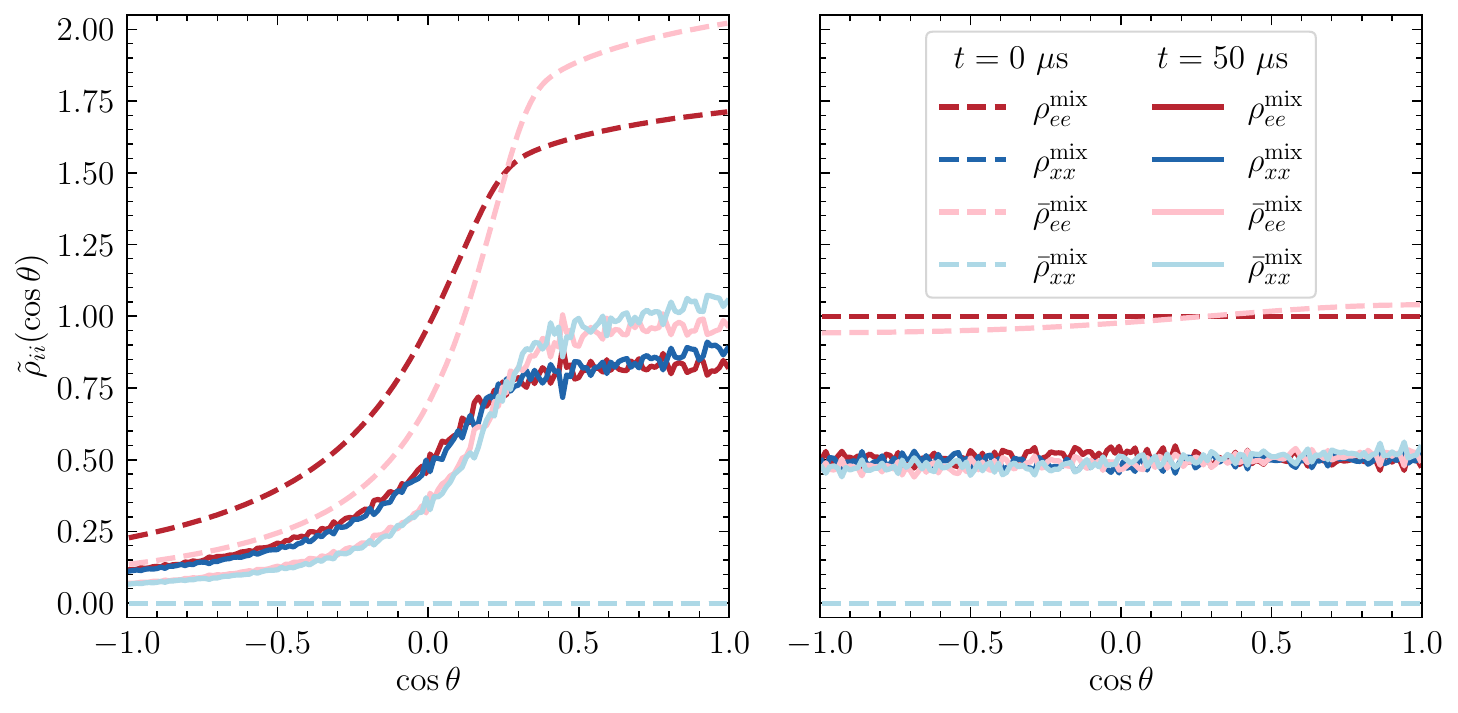}
\caption{Radial average of $\rho_{ii}$, before flavor conversion (dashed lines) and once a quasi-steady state configuration is achieved (solid lines, $t=50~\mu$s), as functions of $\cos\theta$ for the simulation shell with periodic boundaries and Case N1. The left and right panels represent two different configurations for the initial angular distributions (our benchmark, forward peaked one on the left and an isotropic one on the right). For both angular configurations, flavor equipartition is achieved across all angular modes.}
\label{Fig:rad_av_rho_gaussian_N1}
\end{figure*}

References~\cite{Zaizen:2023ihz,Wu:2021uvt,Zaizen:2022cik} find that, in the presence of periodic boundaries, flavor equipartition may be achieved only in one part of the angular distribution with respect to the ELN crossing. In order to explore this feature, we compute the radial-averaged $\rho_{ii}$: 
\begin{equation}
 \tilde{\rho}_{ii}(\cos\theta)= \frac{{\int_{r_{\min}}^{r_{\max}} \rho_{ii}(r,\cos\theta)dr}}{\int dr} 
\end{equation}
and show it in Fig.~\ref{Fig:rad_av_rho_gaussian_N1} for Case N1 and two  initial angular distributions. The left panel represents our benchmark forward peaked angular distribution, while the right panel displays our findings for initially nearly isotropic distributions for all neutrino flavors (see Appendix~\ref{App:other_ang_dist}). The radial-averaged quasi-steady state configurations reflect the initial shapes of the angular distributions; yet, after flavor conversion, flavor equipartition for neutrinos and antineutrinos is achieved, and not only on one side of the ELN crossing. We get similar results for Case N2.

The forward-peaked shape of the benchmark angular distributions for all flavors in the quasi-steady state is the reason why the contour plot shows a significant difference between the number density in the forward and backward direction (see left panel of Fig.~\ref{Fig:pert_periodic_gaussian} and  left panel of Fig.~\ref{Fig:rad_av_rho_gaussian_N1}).
As we see on the left panel of Fig.~\ref{Fig:rad_av_rho_gaussian_N1}, the initial forward-peaked shape of the angular distributions is maintained after flavor conversion, with  flavor equipartition being achieved between $\nu_e$ and $\nu_x$ (and similarly for antineutrinos).

Flavor equipartition can take place  across all angular modes or only on one side of the angular distribution with respect to the ELN crossing according to the value of the asymmetry parameter $\alpha={\bar\rho_{ee}}/{\rho_{ee}}$ for the initial angular distributions~\cite{Wu:2021uvt}. 
Computing the asymmetry parameter,  we find $\alpha\simeq 1$ for both angular configurations shown in Fig.~\ref{Fig:rad_av_rho_gaussian_N1}.
If we compare our results with the ones presented in  Fig.~6 of Ref.~\cite{Wu:2021uvt}, 
we find agreement with their case with $\alpha=1$. 
In fact, since we assume  no $\nu_x$ and $\bar\nu_x$ initially,  the total number density is $\int (\mathrm{ELN}$--$\mathrm{XLN}) ~d\cos\theta \simeq 0$. Our results are thus also in agreement with Ref.~\cite{Zaizen:2022cik} that finds flavor equipartition  across the entire angular range if the total number density of $\mathrm{ELN}$--$\mathrm{XLN}$ is zero.

In order to verify that the lepton number is conserved in our periodic simulation shell, as pointed out in Ref.~\cite{Zaizen:2023ihz}, we calculate the angle- and radial-averaged evolution in time for both angular configurations in Fig.~\ref{Fig:rad_av_rho_gaussian_N1} and confirm that the lepton number is indeed constant for both angular configurations. 
Likewise, we do not find conservation of the  lepton number flux for both angular distribution configurations~\cite{Zaizen:2023ihz}.
Figure~2 of Ref.~\cite{Zaizen:2023ihz}
shows that
flavor equipartition is achieved in their simulation with periodic boundaries to the right of the ELN crossing; this happens because $\alpha=0.9$.

\section{Discussion and outlook}
\label{Sec:discussion}
The flavor configuration induced by fast conversion in dense media remains to be unraveled. Because of the challenges intrinsic to the numerical solution of the quantum kinetic equations of neutrinos, the possibility of forecasting the quasi-steady state flavor configuration without numerically solving the equations of motion is very appealing. In this context, a large body of work~\cite{Grohs:2022fyq,Richers:2021xtf,Richers:2022bkd,Bhattacharyya:2020jpj,Bhattacharyya:2022eed,Xiong:2023vcm,Wu:2021uvt,Martin:2019gxb,Martin:2021xyl,Abbar:2021lmm,Duan:2021woc,Zaizen:2023ihz,Zaizen:2022cik} suggests that flavor equipartition may be a general flavor outcome (at least for a part of the angular distribution with respect to the location of the ELN crossing), and perturbations of the neutrino field may play an important role in speeding up the achievement of flavor equipartition, allowing fast oscillation waves to diffuse in space across spatial scales. 
On the other hand, it has been shown in specific cases that, once the condition of periodic boundaries of the simulation is relaxed, flavor equipartition may be one of the possible outcomes, but not the only one~\cite{Shalgar:2022rjj,Shalgar:2022lvv,Padilla-Gay:2022wck,Nagakura:2022qko,Nagakura:2022kic,Nagakura:2022xwe,Sigl:2021tmj}.

This work aims to carry out a systematic investigation of the expected flavor outcome in two different simulation setups: 1.~a spherically symmetric simulation shell without periodic boundaries, with dynamically evolving angular distributions thanks to non-forward scatterings of neutrinos with the background medium and neutrino advection; 2.~a periodic simulation shell, with arbitrarily fixed angular distributions of neutrinos and  advection.
For each simulation setup, we consider two different configurations for the implementation of the perturbations in the neutrino field, in one case perturbations of different intensity appear randomly along the locations of the ELN crossings  and in the other case perturbations randomly appear across the simulation shell (independent of the ELN loci). We find that flavor equipartition is generally achieved in the simulation shell with periodic boundaries, independent of the initial flavor configuration. However, flavor equipartition is not a common flavor outcome when the simulation shell does not have periodic boundaries. 

Any random fluctuations of the neutrino field in the diagonal and off-diagonal terms of the density matrix allows for the achievement of flavor equipartition in the simulation shell with periodic boundaries.  The periodicity in the simulation shell aids the diffusion of flavor structures towards smaller and smaller scales. On the other hand, the same system, without perturbations, does not achieve flavor equipartition.
Comparing two different initial angular distributions, one that is forward peaked and one that is nearly isotropic, we find that flavor equipartition is achieved in the quasi-steady state for both cases across all angular modes. As suggested by Refs.~\cite{Wu:2021uvt,Zaizen:2022cik}, this is due to the asymmetry parameter between $\nu_e$ and $\bar\nu_e$ being close to $1$ for both our angular configurations. 

On the other hand, we find that random fluctuations of the neutrino field modify the final flavor configuration in the simulation shell without periodic boundaries, but they are pushed out of the simulation shell after a certain time because of neutrino advection, therefore not necessarily leading to flavor equipartition. 

In summary, fast flavor conversion may lead to a very rich flavor phenomenology in neutrino-dense sources, such as core-collapse supernovae and neutron star mergers with implications on the source physics. While it would be helpful to forecast the final flavor configuration, this work shows that the choice of the boundary conditions in the numerical simulation strongly affects the flavor outcome in the presence of flavor conversion.

\acknowledgments
We thank Huaiyu Duan for useful discussions. This project has received support from the Villum Foundation (Project No.~13164), the Danmarks Frie Forskningsfond (Project 
No.~8049-00038B), the European Union (ERC, ANET, Project No.~101087058), and the Deutsche Forschungsgemeinschaft through Sonderforschungbereich SFB 1258 ``Neutrinos and Dark Matter in Astro- and Particle Physics'' (NDM). 
Views and opinions expressed are those of the authors only and do not necessarily reflect those of the European Union or the European Research Council. Neither the European Union nor the granting authority can be held responsible for them.

\appendix
\section{Simulation shell with periodic boundaries and isotropic angular distribution}
\label{App:other_ang_dist}

In this Appendix, we explore the influence of the choice of the initial angular distribution for the quasi-steady state configuration achieved in the presence of periodic boundaries.
To this purpose, we use an isotropic distribution for $\nu_e$ and a nearly isotropic angular distribution for $\bar\nu_e$:
\begin{equation}
 \rho_{ee}(r,\cos{\theta},t=0) = 1
\end{equation}
\begin{equation}
 \bar{\rho}_{ee}(r,\cos{\theta},t=0) = 2 \left[0.47 + 0.05 \exp\left(-(\cos\theta-1)^2\right)\right]\ ;
\end{equation}
this choice led to flavor instabilities in Ref.~\cite{Shalgar:2020wcx}. The distributions are normalized such that $\int \rho_{ee}~ d\cos\theta / \int d\cos\theta = 1$.
We set the distributions for $\nu_x$ and $\bar\nu_x$ to zero for simplicity.

\begin{figure*}
\centerline{Periodic boundaries, with perturbations,  Case N1}
\vspace{0.5cm}
\includegraphics[width=0.99\textwidth]{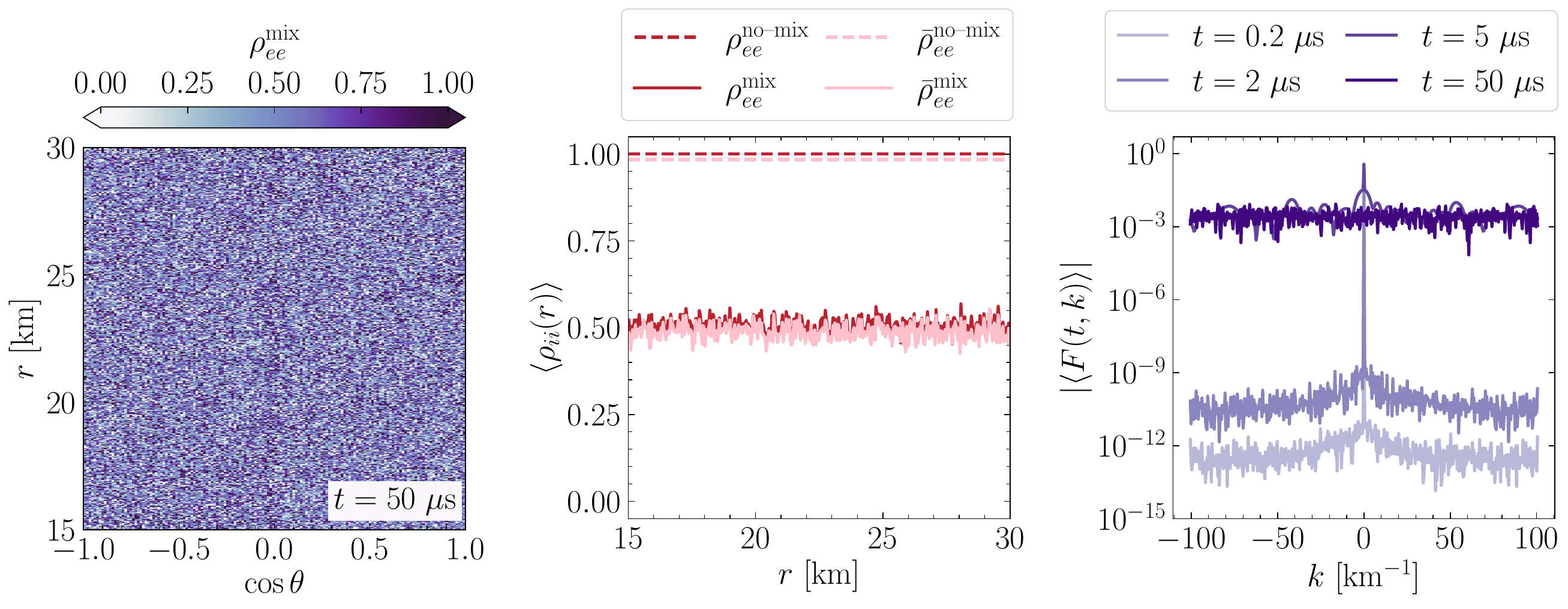}

\vspace{0.5cm}
\centerline{Periodic boundaries, with perturbations, Case N2}

\vspace{0.5cm}
\includegraphics[width=0.99\textwidth]{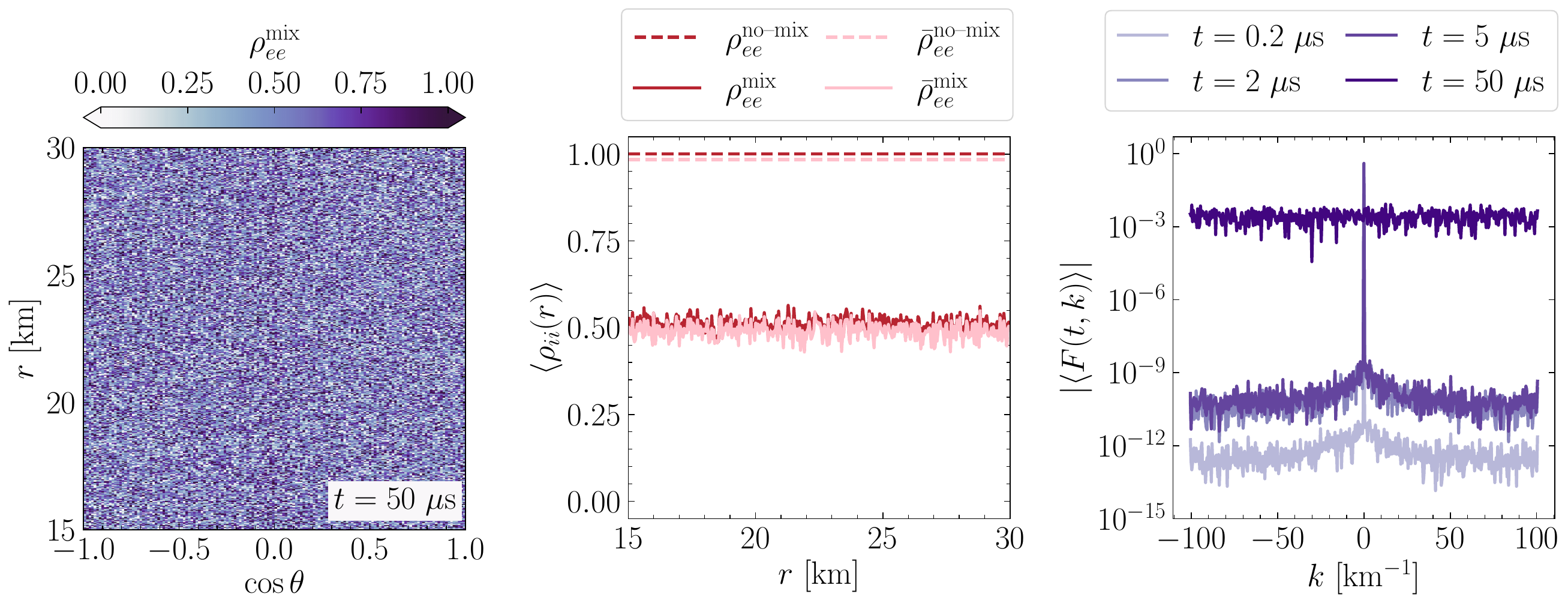}
\caption{Quasi-steady state neutrino configuration with perturbations and periodic boundary conditions, for Cases N1 (top panels) and N2 (bottom panels). {\it Left panels:} Contour plot of $\rho_{ee}$ in the presence of flavor conversion at $t=50~\mu$s in the plane spanned by $\cos\theta$ and $r$. 
{\it Middle panels:} Angle-averaged radial profile of $\rho_{ee}$ (red) and $\bar\rho_{ee}$ (pink) with (solid lines) and without (dashed lines) flavor conversion. 
{\it Right panels:} Angle-averaged Fourier power spectrum of the off-diagonal component of the density matrix $\rho_{ij}$ (see Eq.~\ref{eq:Fourier}) for different time snapshots. For both Cases N1 and N2, flavor conversion spreads across the simulation shell and flavor equipartition is achieved.  Flavor waves cascade down to smaller and smaller scales, until all Fourier modes are uniformly occupied.}
\label{Fig:pert_periodic_other_ang_gaussian}
\end{figure*}
The angular configuration above is implemented in the  simulation shell with periodic boundaries and Gaussian perturbations. The top (bottom) panels of Fig.~\ref{Fig:pert_periodic_other_ang_gaussian} show the results for Case N1 (Case N2). The left panels show that the quasi-steady state flavor configuration has small-scale fluctuations across the radial and angular range. There is thus no difference between the forward and backward moving neutrinos, as found Fig.~\ref{Fig:pert_periodic_gaussian}. 

The middle panels of Fig.~\ref{Fig:pert_periodic_other_ang_gaussian} represent the angle-averaged radial profiles of the density matrices for $\nu_e$ and $\bar\nu_e$. We see that $\rho_{ee}$ and $\bar\rho_{ee}$ move from an initial value of $1$ to  $\simeq 0.5$ once the quasi-steady state configuration is achieved. 

The right panels of Fig.~\ref{Fig:pert_periodic_other_ang_gaussian} show the Fourier power spectrum (see Eq.~\ref{eq:Fourier}) as a function of $k$. While the Fourier power spectrum  peaks at  $k=0$ at the beginning, higher order modes are populated as the system evolves in time. From  $t \simeq 5~\mu$s for Case N1, all Fourier modes are uniformly occupied; the same occurs for  Case N2 at a slightly later time. This is similar to the right panels of Fig.~\ref{Fig:pert_periodic_gaussian}.

\section{Neutrino flavor evolution with radial perturbations}
\label{App:radial_perts}
In addition to the Gaussian perturbations, we also investigate  the impact of radial perturbations.
We perturb the system to obtain a radial dependence in $\rho_{ee}$ 
\begin{equation}
 \delta\rho_{ee}(r,\cos{\theta},t=0) = 10^{-8} \cos\left[\frac{2\pi(r-r_{\min})}{r_{\max}-r_{\min}}\right]\ .
\end{equation}

\begin{figure*}
\centerline{No periodic boundaries, radial perturbations}
\vspace{0.5cm}

\includegraphics[width=0.55\textwidth]{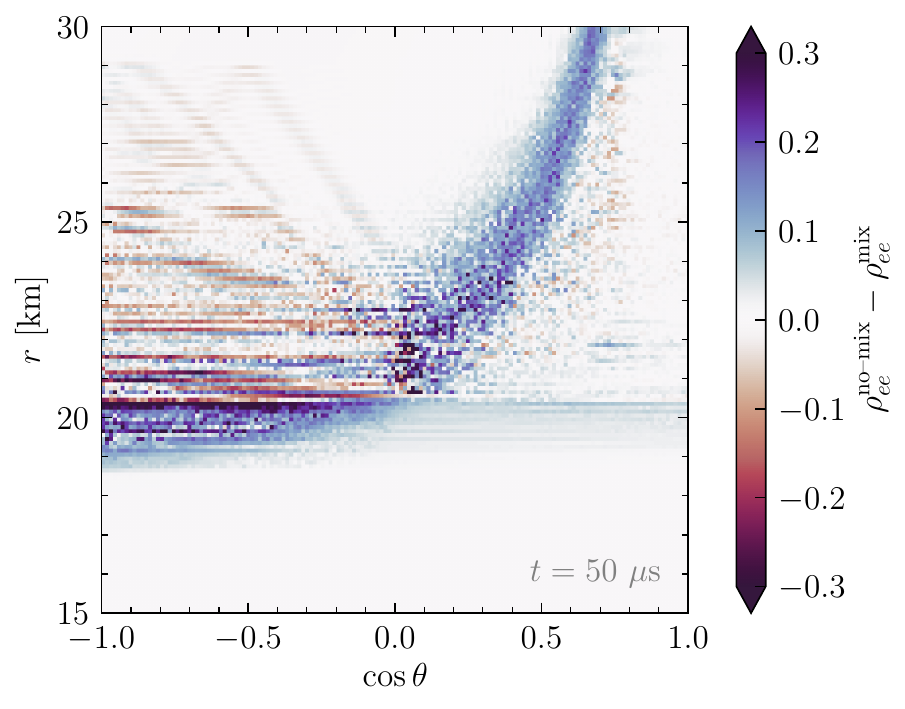}
\includegraphics[width=0.44\textwidth]{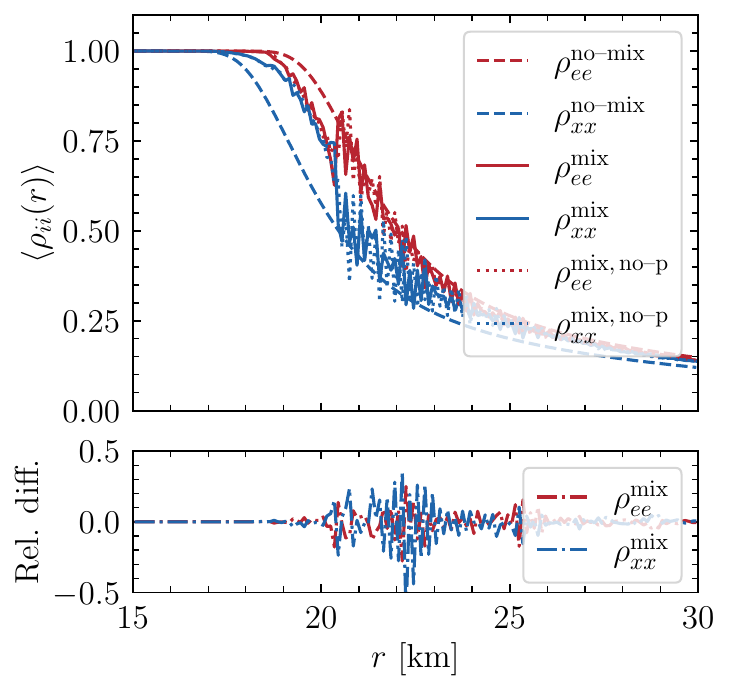}
\caption{Quasi-steady state neutrino configuration with radial perturbations and no periodic boundary conditions. {\it Left panel:} Contour plot of the difference between $\rho_{ee}$ with and without flavor conversion at $50~\mu$s when the quasi-steady state configuration is reached in the plane spanned by $\cos\theta$ and $r$. {\it Right panel:} Radial evolution of the angle-averaged $\rho_{ee}$ (in red) and $\rho_{xx}$ (in blue)  with flavor conversion (solid lines) and without flavor conversion (dashed lines). Also, in this scenario, flavor equipartition is not achieved and the final flavor configuration is minimally affected with respect to the case without perturbations.}
\label{Fig:pert_radial}

\end{figure*}

Figure~\ref{Fig:pert_radial} displays the final flavor configuration obtained for the non-periodic simulation. The left panel shows the quasi-steady state solution at $t=50~\mu$s, which is similar to the one plotted in the middle panel of Fig.~\ref{Fig:no_perts} of the simulation without perturbations. The right panel shows the angle-averaged density profile, which should be compared to the right panel of Fig.~\ref{Fig:no_perts}. As for Cases N1 and N2, we see an average difference of less than $10\%$, with flavor equipartition between $\nu_e$ and $\nu_x$ not being  reached for any radii. We conclude that the final flavor configuration is negligibly affected by the presence of radial perturbations. 

\begin{figure*}
\centerline{Periodic boundaries, radial perturbations, benchmark angular distribution}
\vspace{0.5cm}
\includegraphics[width=0.99\textwidth]{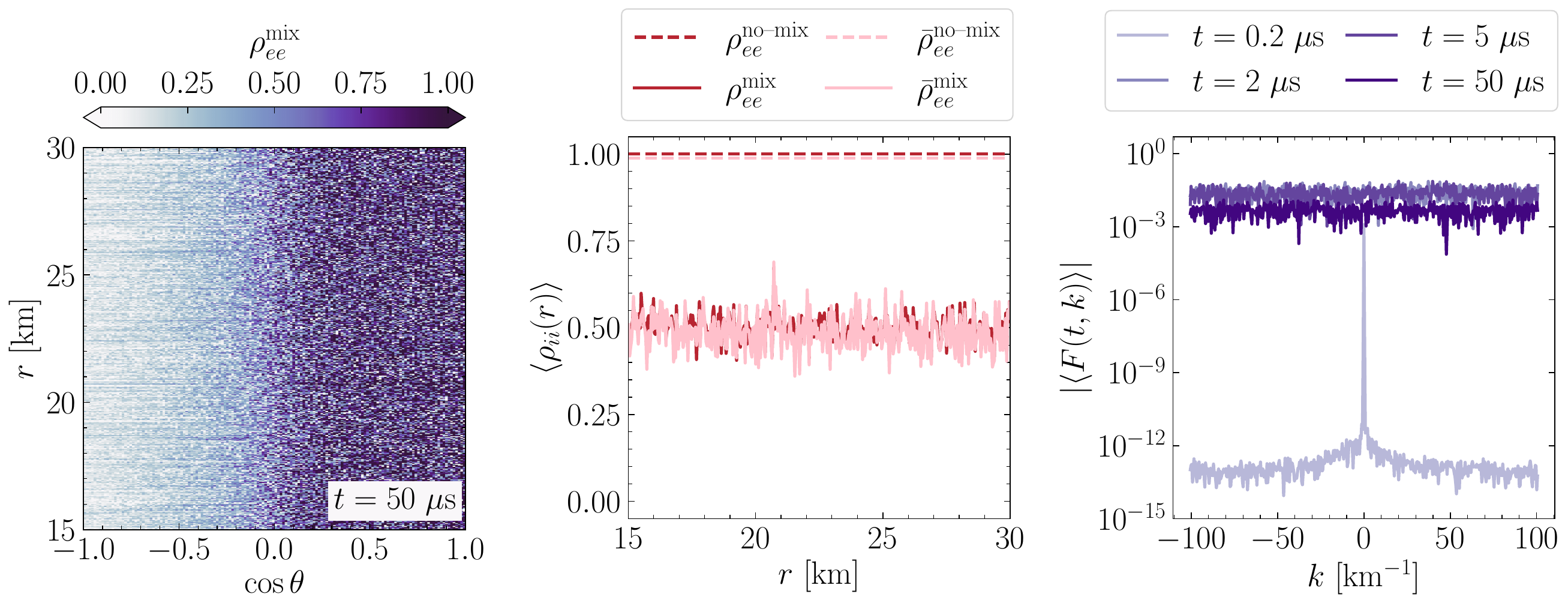}

\vspace{0.5cm}
\centerline{Periodic boundaries, radial perturbations, isotropic angular distribution}

\vspace{0.5cm}
\includegraphics[width=0.99\textwidth]{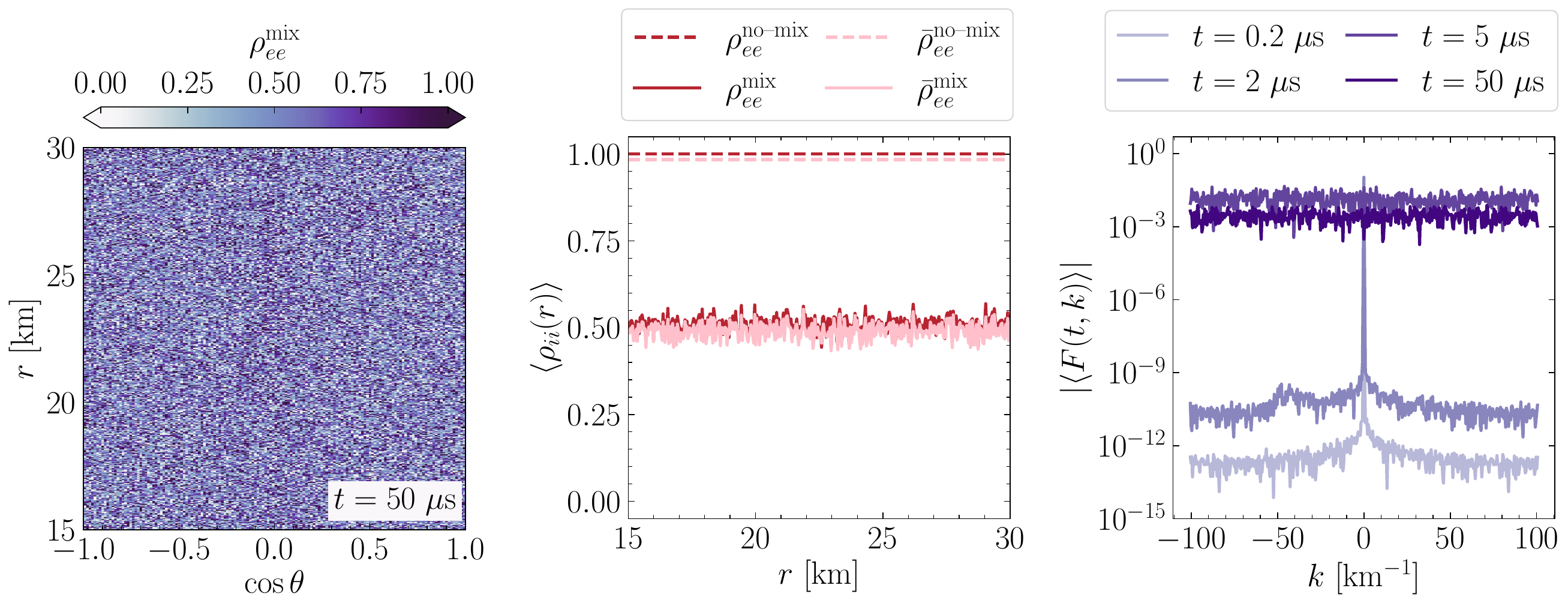}
\caption{Quasi-steady state neutrino configuration with radial perturbations and periodic boundary conditions, for the benchmark angular distributions (top panels) and the isotropic angular distributions (bottom panels). {\it Left panels:} Contour plot of $\rho_{ee}$ in the presence of flavor conversion at $t=50~\mu$s in the plane spanned by $\cos\theta$ and $r$. 
{\it Middle panels:} Angle-averaged radial profile of $\rho_{ee}$ (red) and $\bar\rho_{ee}$ (pink) with (solid lines) and without (dashed lines) flavor conversion. 
{\it Right panels:} Angle averaged Fourier power spectrum of the off-diagonal component of the density matrix $\rho_{ij}$ (see Eq.~\ref{eq:Fourier}) for different time snapshots. 
For both Cases N1 and N2, flavor conversion spreads across the simulation shell and flavor equipartition is achieved and all Fourier modes are uniformly occupied.}
\label{Fig:pert_periodic_radial}
\end{figure*}

Figure~\ref{Fig:pert_periodic_radial} displays the final flavor configuration obtained after introducing  radial perturbations in the simulation shell  with periodic boundaries. The top panels show the quasi-steady state configuration for our benchmark angular distribution, while the bottom panels show the isotropic angular distribution. The top left panel shows a difference in the neutrino number density between the forward and backward direction which is a reflection of the initial angular distributions. For both angular configurations, we obtain flavor equipartition when averaging over angle, shown by the radial profiles in the middle panels. The right panels show the Fourier power spectrum. For the benchmark angular distribution, all Fourier modes become uniformly occupied  at $t \simeq 2~\mu$s, whereas this happens a bit later for the isotropic angular distribution.

\bibliographystyle{JHEP}
\bibliography{references}

\end{document}